\pdfoutput=1

\documentclass[11pt,twoside,a4paper,cmspaper,final,collab]{cms-tdr}
\def\svnVersion{174611}\def\cmsCernNo{2012-224}\def\cmsCernDate{\today}\def\cmsMessage{Submitted to the Journal of High Energy Physics}

\begin{document}\cmsNoteHeader{TOP-11-005}

\hyphenation{had-ron-i-za-tion}
\hyphenation{cal-or-i-me-ter}
\hyphenation{de-vices}
\RCS$HeadURL: svn+ssh://svn.cern.ch/reps/tdr2/papers/TOP-11-005/trunk/TOP-11-005.tex $
\RCS$Id: TOP-11-005.tex 152181 2012-10-11 09:09:33Z jandrea $

\providecommand{\POWHEG} {\textsc{Powheg}\xspace}

\def\mrm{\mathrm}
\def\ra{\rightarrow}

\newcommand{\roots}{\ensuremath{\sqrt{s}}}
\newcommand{\lhcE}[1]{\ensuremath{\roots ={#1}~\TeV}}

\newcommand{\FIXMEC}[1]{{\textcolor{red}{#1}}}

\renewcommand{\sign}{\ensuremath{\mathrm{sign}}}
\newcommand{\rphi}{\text{$r$-$\phi$}}
\newcommand{\etaphi}{\text{$\eta$-$\phi$}}
\newcommand{\rz}{\text{$r$-$z$}}
\newcommand{\met} {\ensuremath{E\!\!\!\!/_{\mathrm{T}}}}
\newcommand{\jpt}{\ensuremath{\mathrm{JPT}}}
\newcommand{\tcmet}{\ensuremath{\mrm{tcMET}}}
\newcommand{\calomet}{\ensuremath{\mrm{\met^{\mrm{calo,Type1}}}}}
\newcommand{\pfmet}{\ensuremath{\mrm{\met^{\mrm{PF}}}}}
\newcommand{\jet}{\ensuremath{\mrm{jet}}}
\newcommand{\jets}{\ensuremath{\mrm{jets}}}
\newcommand{\njet}{\ensuremath{N_\jet}}
\newcommand{\isotk}{\ensuremath{I_\mrm{trk}}}
\newcommand{\isocal}{\ensuremath{I_\mrm{cal}}}
\newcommand{\isocomb}{\ensuremath{I_\mrm{rel}}}
\newcommand{\mll}{\ensuremath{M_{\ell\ell}}}
\newcommand{\etsc}{\ensuremath{E_{T}^{\mrm{sc}}}}
\newcommand{\dxybs}{\ensuremath{d_0^{\mrm{BS}}}}

\newcommand{\PZ}{\ensuremath{\cmsSymbolFace{Z}}}
\newcommand{\jp}{\ensuremath{\PJgy}}
\newcommand{\eepm}{\ensuremath{\Pep\Pem}}
\newcommand{\mmpm}{\ensuremath{\Pgmp \Pgmm}}
\newcommand{\ttpm}{\ensuremath{\Pgt^+ \Pgt^-}}
\newcommand{\empm}{\ensuremath{\Pe^\pm \Pgm^\mp}}
\newcommand{\dy}{\ensuremath{\PZ/\Pgg^\star}}
\newcommand{\dyee}{\ensuremath{\dy\to\eepm}}
\newcommand{\dymm}{\ensuremath{\dy\to\mmpm}}
\newcommand{\dytt}{\ensuremath{\dy\to\ttpm}}
\newcommand{\Zee}{\ensuremath{\PZ\to\eepm}}
\newcommand{\Zmm}{\ensuremath{\PZ\to\mmpm}}
\newcommand{\wen}{\ensuremath{\PW\to \Pe\Pgne}}
\newcommand{\wmn}{\ensuremath{\PW\to\Pgm\Pgngm}}
\newcommand{\wtn}{\ensuremath{\PW\to\Pgt\Pgngt}}
\newcommand{\tW}{\ensuremath{\cmsSymbolFace{t}\PW}}
\newcommand{\VV}{\ensuremath{\cmsSymbolFace{VV}}}
\newcommand{\WoZ}{\ensuremath{\PW/\PZ}}

\newcommand{\EEE}[1]{\ensuremath{\times 10^{#1}}}
\newcommand{\pp}{\ensuremath{\Pp\Pp}}
\newcommand{\ppbar}{\ensuremath{\Pp\Pap}}

\newcommand{\totLumi}{2.3\fbinv}
\newcommand{\totLumiWerr}{\ensuremath{2.3\pm 0.1\fbinv}}

\newcommand{\nNoNo}{\ensuremath{N_{\overline{n}\overline{n}}}}
\newcommand{\nNoNu}{\ensuremath{N_{{n}\overline{n}}}}
\newcommand{\nNuNu}{\ensuremath{N_{{n}{n}}}}

\newcommand{\sye}[1]{\ensuremath{~\pm #1}}
\newcommand{\ase}[2]{\ensuremath{_{~- #1}^{~+ #2}}}
\newcommand{\asi}[2]{\ensuremath{_{~- #1}^{~+ #2}}}

\newcommand{\Routin}{\ensuremath{R_\mrm{F/P}}}
\newcommand{\DRoutin}{\ensuremath{\delta(\Routin)/\Routin}}
\newcommand{\RTL}{\ensuremath{R_\mrm{TL}}}

\newcommand{\ns}{\ensuremath{\mrm{ns}}}

\cmsNoteHeader{TOP-11-005}
\title{Measurement of the \ttbar\ production cross section in the dilepton channel in pp collisions at $\sqrt{s} =7$\TeV}

\date{\today}

\abstract{
The $\ttbar$ production cross section ($\sigma_{\ttbar}$) is measured in proton-proton collisions at $\sqrt{s} = 7\TeV$
in data collected by the CMS experiment, corresponding to an integrated luminosity of 2.3\fbinv.
 The measurement is performed in events with two
leptons (electrons or muons) in the final state, at least two jets identified as jets originating from b quarks,
and the presence of an imbalance in transverse momentum. The measured value
of $\sigma_{\ttbar}$ for a top-quark mass of 172.5\GeV is $161.9\pm2.5\,\text{(stat.)}\,{}^{+5.1}_{-5.0}\,\text{(syst.)}\pm3.6\,\text{(lumi.)}\unit{pb}$,
consistent with the prediction of the standard model.
}

\hypersetup{%
pdfauthor={CMS Collaboration},%
pdftitle={Measurement of the t t-bar production cross section  in the dilepton channel in pp collisions at sqrt(s)=7 TeV},%
pdfsubject={CMS},%
pdfkeywords={CMS, physics, software, computing}}

\maketitle %maketitle comes after all the front information has been supplied
\newcommand{\mymet}{\makebox[2.4ex]{\ensuremath{\not\!\! E_{\mathrm{T}}}}}
\newcommand{\myttbar}{\ensuremath{{t\overline{t}}}\xspace}
\newcommand{\Bot}{\ensuremath{\cmsSymbolFace{b}}\xspace}

\newcommand{\usedLumiUp}{\ensuremath{2.3\pm0.1\fbinv}}
\newcommand{\usedLumi}{\ensuremath{2.3\pm0.1\fbinv}}

\providecommand{\eepm}{\ensuremath{\Pep\Pem}}               % exists in pdefs :(
\providecommand{\mmpm}{\ensuremath{\Pgmp \Pgmm}}            % exists in pdefs :(
\providecommand{\ttpm}{\ensuremath{\Pgt^+ \Pgt^-}}          % exists in pdefs :(
\providecommand{\empm}{\ensuremath{\Pe^\pm \Pgm^\mp}}       % exists in pdefs :(
\providecommand{\pp}{\ensuremath{\Pp\Pp}}                   % exists in pdefs :(
\providecommand{\ppbar}{\ensuremath{\Pp\Pap}}               % exists in pdefs :(
\providecommand{\ase}[2]{\ensuremath{_{~- #1}^{~+ #2}}}     % exists in pdefs :(
\providecommand{\roots}{\ensuremath{\sqrt{s}}}              % exists in pdefs :(
\providecommand{\lhcE}[1]{\ensuremath{\roots ={#1}~\TeV}}   % exists in pdefs :(
\providecommand{\dy}{\ensuremath{\PZ/\Pgg^\star}}           % exists in pdefs :(
\providecommand{\dyee}{\ensuremath{\dy\to\eepm}}            % exists in pdefs :(
\providecommand{\dymm}{\ensuremath{\dy\to\mmpm}}            % exists in pdefs :(
\providecommand{\dytt}{\ensuremath{\dy\to\ttpm}}            % exists in pdefs :(
\providecommand{\mll}{\ensuremath{M_{\ell\ell}}}            % exists in pdefs :(
\def\mrm{\mathrm}                                       % exists in pdefs :(
\providecommand{\isocomb}{\ensuremath{I_\mrm{comb}}}    % other definition in pdefs.tex???
\providecommand{\WoZ}{\ensuremath{\PW/\PZ}}               % exists in pdefs :(
\providecommand{\tW}{\cPqt\PW}        % other definition in pdefs.tex???
\providecommand{\VV}{\ensuremath{\mathrm{VV}}}          % other definition in pdefs.tex???
\providecommand{\ns}{\ensuremath{\mrm{ns}}}               % exists in pdefs :(
\providecommand{\wmn}{\ensuremath{\PW\to\Pgm\Pgngm}}      % exists in pdefs :(

\newcommand{\ROutIn}{$R_\text{out/in}$}

\section{Introduction}
\label{sec:introduction}
A precise measurement of the \ttbar\ production cross section ($\sigma_{\ttbar}$) in pp collisions
 at the Large Hadron Collider (LHC) is important for several reasons.
 At LHC energies, \ttbar\ production is dominated by gluon-gluon fusion,
 representing a benchmark for other processes of the standard model (SM) initiated through the same mechanism,
 such as the production of Higgs bosons.
 In addition, a precise value of $\sigma_{\ttbar}$ can provide constraints on parton distribution functions (PDF)
and be used to check the validity of perturbative calculations in quantum chromodynamics (QCD).
It can also benefit searches for physics beyond the SM, as \ttbar\ production is often a major source of background.

The precision of initial measurements of $\sigma_{\ttbar}$ in 2010
at $\sqrt{s}$ = 7\TeV~\cite{top10001,ATLAStopPublication4} has been
improved in Refs.~\cite{ATLAStopPublication5,ATLAStopPublication6,CMStopPublication2,CMStopPublication3,CMStopPublication5,ATLASttWPublication},
and measurements of electroweak single-top-quark
production are available in Refs.~\cite{CMStopPublication4,ATLASttchenlPublication}.
By now, many top-quark events have been collected at the LHC, with studies proceeding on a variety of top-quark production and decay
channels~\cite{CMStopPublication2,CMStopPublication6,ATLAStopPublication1,ATLAStopPublication2},
as well as on searches for deviations relative to predictions of the SM.

This work presents a measurement of $\sigma_{\ttbar}$ in dilepton final states
that improves upon a previous measurement~\cite{CMStopPublication2} of the Compact Muon Solenoid (CMS) experiment~\cite{JINST}. It is based on more refined event
selections and analysis techniques, an improved estimation of systematic uncertainties, and with a data sample approximately sixty times larger.
Starting from the nearly 100\% decay of both top quarks through the electroweak transition ${\cPqt\rightarrow \PW \cPqb}$,
we focus on events in the dilepton final states \eepm, \mmpm, and \empm, where both \PW\ bosons decay leptonically
($\PW\rightarrow \ell \nu_{\ell}$),
 but with contributions from $\PW\rightarrow$ $\tau \nu_{\tau}$ arising only when the $\tau$ leptons decay into $\ell \nu_{\rm \ell}\nu_{\tau}$ ($\ell=$ e or $\mu$) states.
 A cross-section measurement based on $\tau$ leptons that decay into hadrons and a neutrino can be found in Ref.~\cite{Xsec_taus}.
The final states contain two leptons of opposite electric charge, their accompanying neutrinos from
the \PW-boson decays, and two jets of particles resulting from the hadronization of the \Bot\ quarks.
These modes correspond to ($6.45 \pm 0.11$)\%~\cite{PDG2012} of all \ttbar\ decay channels, including
the partial contributions from $\tau$ leptons.% leptons that subsequently decay to electrons and muons, as it is done here.

This measurement is based on a data sample corresponding to an integrated luminosity of \totLumiWerr\
recorded by the CMS experiment.
 A profile likelihood-ratio (PLR) method~\cite{plr1, plr2},
as well as a counting analysis based on
direct selections, %(cutoffs) on kinematic and geometric variables,
are used to extract the \ttbar cross section.

A brief description of the components of the CMS detector specific to this analysis is provided in Section~\ref{sec:detector},
followed by details of event Monte Carlo (MC) simulation in Section~\ref{sec:simulation}, and event selection in Section~\ref{sec:event_selection}.
Estimations of backgrounds yields are presented in Section~\ref{sec:bkgd}.
Following a discussion of systematic uncertainties in Section~\ref{sec:systematics},
the event yields in data are compared with simulations in Section~\ref{sec:yields}.
The results of the measurement of the \ttbar\ cross section are presented in Section~\ref{sec:measure},
 followed by a summary in Section~\ref{sec:conclusions}.

\section{The CMS detector}
\label{sec:detector}

The central feature of the CMS apparatus is a superconducting solenoid,
13\unit{m} in length and 6\unit{m} in diameter, which provides an axial magnetic
field of 3.8\unit{T}.  The bore of the solenoid is outfitted with a variety of
particle-detection systems.  Charged-particle trajectories are
measured with the silicon pixel and strip trackers that cover $0 < \phi <
2\pi$ in azimuth and $|\eta |<2.5$ in pseudorapidity, where $\eta$ is
defined as $\eta =-\ln[\tan(\theta/2)]$, with $\theta$ being the
polar angle of the trajectory of the particle with respect to the
anticlockwise beam direction.
A crystal electromagnetic calorimeter
(ECAL) and a brass/scintillator hadronic calorimeter surround
the inner tracking volume
and  provide high-resolution measurements of energy ($E$) used to reconstruct electrons, photons and particle jets.
Muons are measured in
gas detectors embedded in the flux-return yoke of the solenoid.
The detector is nearly hermetic, thereby providing reliable
measurements of momentum imbalance in the plane transverse to the beams.
A two-tier trigger system selects the most interesting \pp\ collisions for analysis.
A more detailed description of
the CMS detector is given in Ref.~\cite{JINST}.

\section{Simulation of signal and backgrounds}
\label{sec:simulation}
The \ttbar\ cross section, calculated
with the {\sc mcfm} program~\cite{mcfm,mcfm:tt} at next-to-leading order (NLO) in perturbative QCD,
assuming a top-quark mass of $m_{\cPqt}= 172.5\GeV$, is $158^{+23}_{-24}\unit{pb}$.
Several approximate next-to-next-to-leading-order (NNLO) calculations are also available in the literature, in particular those of
Kidonakis ($163\ase{10}{11}\unit{pb}$~\cite{kidonakis:2010dk}), Ahrens et al. ($149\pm11\unit{pb}$~\cite{Ahrens:2010zv})
and the \textsc{HATHOR} program ($164\ase{10}{13}\unit{pb}$~\cite{Langenfeld:20011}), with the latter
used to normalize the distributions and yields of simulated \ttbar\ events to the measured luminosity.
However, this cross section is used just to present expected rates in figures and tables and has no effect on the final measurement.

Signal and background events are simulated using the MC event generators
\MADGRAPH~(v. 5.1.1.0)~\cite{madgraph}, \POWHEG~(r1380)~\cite{powheg} or \PYTHIA~(v. 6.424)~\cite{pythia}, depending on the process in question.
\MADGRAPH is used to model \ttbar\ events, with matrix elements corresponding to up to three additional partons that are matched to
\PYTHIA, which is used for the subsequent hadronization.
Decays of $\tau$ leptons are handled with \TAUOLA~(v. 2.75)~\cite{tauola}.

\MADGRAPH is also used to simulate the
\PW+jets and Drell-Yan (DY: $\cPq\cPaq\rightarrow \dy$+jets, where $\Pgg^\star$ represents the contribution from virtual photons)
background samples, with up to 4 partons in the final state.
For \PW+jets, the \MADGRAPH sample provides only event with leptonic decays of \PW\ bosons.
The total inclusive NNLO cross section of  $31.3\pm 1.6\unit{nb}$
is calculated with the \textsc{fewz} program~\cite{fewz}.
Simulations of DY events with two oppositely charged leptons of same flavor in the final state are generated for
dilepton invariant masses between 10 and 50\GeV, as well as for ${>}50\GeV$. The corresponding NNLO cross sections,
also calculated with \textsc{fewz} are  $11.91\pm 0.64\unit{nb}$ and $3.04\pm 0.13\unit{nb}$, respectively.
Single-top-quark  events ($\pp\to \mathrm{tW^-}$ and $\pp\to\mathrm{\bar{t}W^+}$)
are simulated in \POWHEG, for a cross section
of $15.7\pm 0.4\unit{pb}$ (calculated at approximate NNLO~\cite{kidonakis}).
Inclusive production of the \PW\PW, \PW\PZ, and \PZ\PZ\, diboson final states is simulated with \PYTHIA,
 and the respective cross sections are $47.0\pm1.5\unit{pb}$, $18.2\pm 0.7\unit{pb}$, and $7.7\pm 0.2\unit{pb}$.

Among all of the simulated backgrounds, only the contributions from
\PW\PW, \PW\PZ, and \PZ\PZ\ (referred as diboson in the text and labeled as \VV\ in figures)
and single-top-quark production are used to estimate the absolute number of background events.
Recent measurements of single-top-quark and diboson cross sections show good agreement with SM
predictions~\cite{CMStopPublication4,ATLASttWPublication,CMStopPublication7,ATLASWWPublication,ATLASWZPublication,ATLASZZPublication}.
All other backgrounds are estimated from control samples in data.

Effects from additional pp interactions (pileup) are modeled by adding simulated minimum-bias events (generated with \PYTHIA) to the simulated processes,
using a pileup multiplicity distribution that reflects
the luminosity profile of the observed pp collisions. The CMS detector response is simulated using \GEANTfour (v.~9.4)~\cite{geant}.

\section{Event selection}
\label{sec:event_selection}
Events are collected using dilepton triggers (excluding ``hadronic" $\tau$ triggers) that require the presence of
two large transverse momentum ($\pt$) leptons.
The definitions and ranges of thresholds, isolation and identification requirements for dilepton triggers were changed periodically during data taking so as to adapt to changes
in instantaneous luminosity delivered by the LHC.
In particular, electrons selected at the trigger stage must pass a threshold on electron transverse energy $\ET$, that ranges from 8 to 17\GeV, as measured online by combining
information from ECAL with that from the inner tracker, and
muons identified in muon detectors are similarly required to exceed a threshold on $\pt$, that ranges between 7 and 17\GeV
as measured online by combining information from the outer muon detector with that from the inner tracker.

The efficiency for \ttbar\ dilepton triggers is measured in data
through other triggers
that are only weakly correlated with the dilepton-trigger requirements.
Because of the presence of a significant imbalance in transverse momentum in \ttbar\ events,
\met-based triggers with different \met\ thresholds are used for this purpose.
At the trigger level, the \met\ is defined by the magnitude of the vector $\vec{\met} = - \sum{ \vec{E}_\mathrm{T}}$,
using the transverse energies of calorimeter towers.
 Studies based on simulated events indicate that the \met\ triggers are
 uncorrelated with the dilepton triggers.

Using the measured dilepton-trigger efficiency in data,
the corresponding efficiencies in simulations are corrected by multiplicative data-to-simulation scale factors (SF) of
$0.962\pm 0.015$, $0.977 \pm 0.016$, and $1.008 \pm 0.009$
for the \eepm, \mmpm\ and \empm\ final states, respectively, to provide agreement between data and simulation.

The first step in the offline selections requires the presence of a
proton-proton interaction vertex~\cite{trkpas}
within 24\unit{cm} of the detector center along the beam line direction,
and within 2\unit{cm} of the beam line in the transverse plan.
These selections have an efficiency ${>}99.5\%$ for events with two leptons that pass selection criteria. The main primary vertex of an event
corresponds to the vertex with the largest value for the scalar sum of the transverse momenta of the associated tracks.

At least two leptons in the event (either electrons or muons) are required to pass identification and isolation requirements.
 The selection criteria are similar to those of Refs.~\cite{top10001,CMStopPublication2}, and can be described briefly as follows. Electron candidates~\cite{EGMPAS} are based on
 a cluster of energy depositions in the ECAL. Clusters are
matched to hits in the silicon tracker using a track reconstruction algorithm
that takes into account possible energy loss due to bremsstrahlung.
Muon candidates are reconstructed by combining information from the inner tracker with information from the outer muon detectors~\cite{MUOPAS}.

Both leptons must have $\pt>20\GeV$, with electrons and muons being restricted to $|\eta|<2.5$ and $|\eta|<2.1$, respectively.
The lepton-candidate tracks are required to be consistent with originating
 from the primary vertex, and must satisfy additional quality requirements, as described in Ref.~\cite{top10001,CMStopPublication2}.
Lepton candidates are required to be isolated from other energy depositions in the event.
A cone of $\Delta R = \sqrt{(\Delta \eta)^2 + (\Delta \phi)^2}<0.3$, where
$\Delta \eta$ and $\Delta \phi$ are the differences in pseudorapidity and azimuthal angle between any element of energy and the axis of the lepton, is constructed
around the initial direction of the candidate.
The particle energies within this cone, obtained using the particle-flow (PF) reconstruction algorithm~\cite{PFPAS},
which provides a list of particles and their kinematic properties, are projected onto the plane transverse to the beam, and summed as scalar quantities, excluding the contribution from the lepton candidate.
In this procedure, all charged PF particles not associated with the
main primary vertex are assumed to arise from pileup events, and are excluded from the sum.
Then, the relative isolation discriminant, $I_{\text{rel}}$, is defined as the ratio of this sum to the transverse momentum of the lepton candidate.
A lepton candidate is not considered as isolated and is rejected if the value of $I_{\text{rel}}$ is ${>}0.20$ for a muon and ${>}0.17$ for an
electron. These selections are optimized using simulated \ttbar\  events.

The efficiency of lepton selection is measured using a ``tag-and-probe'' method
in dilepton events enriched  in Z-boson candidates, and indicates that electron and muon reconstruction efficiencies
 are ${>}99\%$~\cite{wzPAS2010}.
The efficiencies for the above lepton-identification requirements, calculated as function of the $\pt$ and $\eta$ of the leptons,
are in the range of 98--99\% for muons and 85--94\% for electrons.
The average efficiency of the lepton-isolation criterion measured in such \PZ\ events
is in the range of 90--99\% for both muons and electrons. The efficiencies measured in data are found to be very close to the
 estimates from the DY simulation.
 Based on an overall comparison of lepton-selection efficiencies in data and simulations,
the event yield in simulation is corrected by multiplicative scale factors of
$0.995\pm 0.003$, $0.997\pm0.005$ and $0.994\pm 0.005$
for the  \eepm, \mmpm\ and \empm\ final states, respectively, to provide consistency with data.

Accepted events are required to have at least one pair of oppositely charged leptons. While muon-charge misidentification is
negligibly small, the electron-charge misidentification rate is 0.8\%, as measured from $\cPZ\to\Pep\Pem$ decays.
 Dilepton \ttbar-candidate events with invariant mass $\mll < 20\GeV$ are removed from all three channels, with a consequent reduction of about 2\% in \ttbar\ signal. However, this requirement significantly suppresses backgrounds from heavy-flavor
resonances, as well as contributions from low-mass  DY processes.
In events containing several possible $\ell^+ \ell^-$ pairs that pass all acceptance requirements,
only the pair of leptons with the highest sum of scalar transverse momenta is retained.
To suppress contributions from \PZ\ boson production, the
invariant mass of the dilepton system is required to be
outside the range of 76 to 106\GeV for both the \eepm\ and \mmpm\ modes.
According to simulation, this invariant mass requirement rejects about 91\% of the DY events,
at the cost of rejecting ${\approx}24\%$ of both the \eepm and \mmpm \ttbar\ signal events.

The anti-$k_\mathrm{T}$ clustering algorithm~\cite{antikt} with a distance parameter $R=0.5$ is used for jet clustering.
Jets are reconstructed based on information from the calorimeter, tracker, and muon systems~\cite{JETPAS}
using PF reconstruction.
Jet energy corrections rely on simulations and on studies performed with exclusive two-jet and photon+jet events in data.
To minimize the impact of pileup, charged PF particles not
associated with the primary event vertex \cite{PFPAS} are ignored in reconstructing jets.
After eliminating the charged component of pileup events, the neutral component is removed by applying a
residual energy correction, following the ``area-based'' procedure described in Refs.~\cite{fastjet1, fastjet2}.
Jets are also required to satisfy $\pt>30\GeV$ and $|\eta|<2.5$.
For dilepton \ttbar\ candidates with at least two jets and before applying any b-tagging requirement, both \Bot-quark jets from \ttbar\ decays pass the jet selection criteria in about 65\% of
the events, as predicted by the \ttbar\ simulation.

For the offline analysis, the missing transverse energy (\met) is redefined by the magnitude of the vectorial sum of the particle transverse momenta
$\vec{\met} = - \sum{ \vec{p}_\mathrm{T}}$,
calculated using the particles reconstructed with the PF algorithm~\cite{METPAS2}.
As implied above, \met\ corresponds to a distinguishing feature of \ttbar\ events in the dilepton channel because of escaping energetic neutrinos.
Neither the dominant DY background (\dyee\ and \mmpm), nor the smaller
background from false isolated leptons in  multijet
events, provide a source of large $\met$.
We require $\met>40\GeV$ in the \eepm\ and \mmpm\ modes
in \ttbar\ events with at least two jets. This cut causes a loss of $\approx$25\% of \ttbar\ signal in these channels, but reduces the
contribution from DY production by about a factor of 30.
 No \met\ requirement is imposed for the \empm\ mode, as there is very little contamination from DY events in this channel.

To account for any mismodeling of the \met\ distribution, the efficiency of the \met\ selection in the \ttbar\ simulation is corrected using data.
The corresponding data-to-simulation SF  are estimated by applying the \met\ selection to  \empm\ data, after
correcting for the presence of background.
The systematic uncertainties on these factors arise from two sources:
(i) the background contamination in the \empm\  channel, which is changed in the study by $\pm$ 30\%, and (ii) the difference in lepton
energy resolution between
electrons and muons, which affects differently the \met\ in \eepm\ and \mmpm\ channels relative to the \empm\ channel.
The data-to-simulation SF are
$1.008\pm 0.012$ and $1.008\pm 0.016$ for the \eepm\ and \mmpm\ final states, respectively,
with the uncertainties accounting for both statistical and systematic effects summed in quadrature.

As dilepton \ttbar\ events contain jets from the hadronization of \Bot\ quarks,
requiring b tagging can
reduce background from events that do not contains b jets.
Jets are identified as  b jets originating from b quarks through a ``combined secondary vertex"
(CSV) algorithm~\cite{BTV-11-004}, which provides a b-tagging discriminant
by combining information from secondary vertices and track-based lifetime measurements.
The chosen b-tagging selection has an efficiency of
80--85\% for each b jet in dilepton \ttbar\
events, and a 10\% mistagging rate for light-flavor or gluon jets misidentified as b jets, both estimated from
inclusive simulated \ttbar\ events.
The \Bot-tagging efficiency can be estimated from \ttbar\ events in data, as described in Ref.~\cite{BTV-11-003}.
To avoid statistical correlations between the extraction of the b-tagging efficiency and the \ttbar\ cross section,
 the b-tagging performance is measured from multijet  events in data \cite{BTV-11-004}, and used to correct
 the b-tagging performance of the simulation. This analysis uses the b-tagging information either as input to a likelihood fit
(in the PLR method presented in Section \ref{sec:results_channels_plr}) or directly to select events that are required to contain at least one b-tagged jet
(in a counting analysis used as a cross-check, and presented in Section \ref{sec:results_channels_counting}).

\section{Background estimates}
\label{sec:bkgd}
The main backgrounds in this analysis arise from DY, diboson, and single-top-quark events (tW), where at least two
prompt leptons are produced from Z or W decays. Other background sources, such as \ttbar\ events with decays to lepton+jets or no leptons at all (all jets), or generic
multijet (MJ) events, are related to the presence of at least one jet reconstructed incorrectly as a lepton, which mainly happens for electrons, or a lepton from the decay of bottom or charm
hadrons, which mainly happens for muons.

The yields from background processes with smallest contributions to \ttbar\ candidates,
corresponding to single-top-quark and diboson production, are estimated directly from simulations.
 This section focuses on other
backgrounds, which are not adequately described through simulation, such as the DY contribution,
and events with leptons that
do not arise from \PW\ or \PZ\ decays. In these cases,
the background estimates are based on data.

\subsection{Backgrounds from Drell-Yan contribution}
\label{sec:dydata}
To estimate the background from DY events contributing into \eepm\ and \mmpm\ \ttbar\ final states, we use the method described in Refs.~\cite{top10001,CMStopPublication2}.
The number of DY events in data
that pass the \PZ-boson veto on $\mll$ can be estimated from
the number of events in data with a dilepton invariant mass within $76< \mll<106\GeV$,
scaled by the ratio (\ROutIn) of events that fail and that pass this selection, which is estimated through DY simulation.

To achieve a better estimate of this background, the value of \ROutIn\ is corrected using data.
In particular, \ROutIn\ is sensitive to detector effects, such as the modeling of lepton resolution and
 the dependence of the fraction of vetoed DY events for different requirements on jet multiplicities.
The value of \ROutIn\ is corrected for these effects by using control regions enriched in DY events in data.

A 15\% systematic uncertainty is assigned
to account for remaining discrepancies in the dependence of \ROutIn\ on \met.
Using the data-corrected \ROutIn\ values leads to
the final data-to-simulation scale factors of $1.86\pm0.50$ and $1.76\pm0.36$
for the \eepm\ and \mmpm\ channels, respectively. After requiring at least one b-tagged jet, there are too few events in data to perform a precise estimation, which leads to larger uncertainties on the corresponding scale factors, yielding $2.04\pm0.52$ and $1.67\pm0.42$ for the \eepm\ and \mmpm\ channels, respectively.

Finally, most of the background contribution in the \empm\ channel is from
$Z/\gamma^* \rightarrow \tau \tau \to \Pe \Pgne \Pgngt { \Pgm \Pgngm \Pgngt}$ events.
As the yield from this final state corresponds to a fraction of about 3\% of the total DY cross section,
 its impact is expected to be small.
In addition, because of the presence of additional neutrinos from $\tau$ decays, the dilepton invariant mass
is often well below the value of the Z-boson mass. The DY contamination of the \empm channel is therefore estimated
through a fit of two components to the dilepton invariant-mass distribution, one component reflecting the dilepton mass distribution for
\dy\ events, and the other corresponding mainly to \ttbar, single-top-quark and diboson events. The templates for these components are extracted from
simulations. The systematic effects that arise
from the small contamination of single-top-quark and diboson events are negligible relative to the large
statistical uncertainty of the fit.
 The data-to-simulation scale factor for the \empm channel, following the jet selection, is found to be $1.34\pm0.29$.

The estimates of the DY contributions to the \ttbar\ data sample are given in Section~\ref{sec:yields},
before and after requiring at least one b-tagged jet.

\subsection{Backgrounds with leptons not from W/Z decays}
\label{sec:fakesdata}
A data-based method is also used to estimate the background from misidentified leptons, and from well-identified
 leptons that pass the isolation requirement but which come from semi-leptonic decays
of bottom or charm hadrons contained within jets.
These leptons are  referred to as non-prompt leptons in the following.
Three categories of backgrounds can be defined: signal-like, W-like, and MJ-like sources,
containing two, one, and no true isolated (prompt) leptons, respectively. The signal-like sample contains  \ttbar\ signal events,
but also DY,  single-top-quark, and diboson events.
The W-like sample consists of W+jets events and \ttbar\ events observed in the lepton+jets channel.
The MJ-like sample contains mainly generic multijet events or \ttbar\ events in all-jets final states.

The number of events in each of the above categories is defined by relaxing the single-lepton isolation from
$I_{\text{rel}}<0.17$ and $I_{\text{rel}}<0.20$ for electrons and muons, respectively (which corresponds to the definition of ``tight'' leptons),
to $I_{\text{rel}}<0.8$ for both flavors of leptons (which defines ``loose'' leptons).
Dilepton samples are then constructed to correspond to loose, medium, and tight events, respectively, defined by two loose,
at least one tight, and two tight isolated leptons.
Only the tight dilepton sample is used to measure $\sigma_{\ttbar}$.

By introducing probabilities for a loose event to pass medium or  tight criteria, and the probability of a medium event to be accepted as a tight
event, a system of equations can be constructed to estimate the number of signal-like, W-like, and MJ-like
events. The probabilities can be expressed in terms of individual probabilities for prompt and non-prompt leptons that pass the relaxed isolation
selection to also pass the tight isolation requirement.
The prompt-lepton efficiencies are estimated using Z events in data, in a manner similar to that discussed in Section~\ref{sec:event_selection},
while the rates for non-prompt leptons are estimated using a data sample enriched in multijet events.
The corresponding systematic uncertainties are estimated by half of the difference between efficiencies determined from data and from simulation studies.

The estimated number of W-like ($N_{\PW}$) and MJ-like ($N_\mathrm{MJ}$) background events in data before and after requiring at least one b-tagged jet are presented
in Table~\ref{tab:fakenobtag}. The uncertainties account for both statistical
and systematic contributions to the efficiency for single prompt and non-prompt loose leptons to pass the tight selection.
The effect of the b-tagging selection efficiency and mistagging rates are estimated from simulation.

\begin{table}[ht]
\begin{center}
\topcaption{Estimated number of W-like ($N_{\PW}$) and MJ-like ($N_\mathrm{MJ}$) background events in data before and after b tagging.}
\begin{tabular}{lccc|ccc}
\hline
\hline
  & \multicolumn{3}{c|} {Before b tagging} & \multicolumn{3}{c} {Requiring $\ge$1 b-tagged jet} \\
\hline
             & \eepm\   & \mmpm\ & \empm\  &
               \eepm\   & \mmpm\ & \empm\ \\
$N_{\rm W}$  &	 $7.8 \pm 5.9$       &    $14.9 \pm 7.1$     &     $63.8 \pm 16.8$  &
             $1.8 \pm 4.8$       &    $9.8 \pm 5.6$     &     $42.4 \pm 14.6$       \\
$N_{\rm MJ}$  &	 $0.7 \pm 0.6$       &    $0.4 \pm 0.3$     &   $21.1 \pm 10.0$     &
             $0.6 \pm 0.5$       &    $0.2 \pm 0.1$     &   $7.5 \pm 3.9$   \\
\hline
\hline
\end{tabular}
\label{tab:fakenobtag}
\end{center}
\end{table}

\section{Sources of systematic uncertainty}
\label{sec:systematics}
Systematic uncertainties considered in this measurement
include those from
biases in detector performance, deviations in \ttbar-signal
acceptance due to ambiguities in modeling \ttbar\ production, precision of background estimates,
and the uncertainty on integrated luminosity~\cite{lumipas}.

A correction is applied to event rates in simulated \ttbar\ samples,
to account for differences in dilepton-trigger and dilepton offline selection efficiencies between data and simulations,
as described in Section~\ref{sec:event_selection}.
The combined data-to-simulation SF,
with their uncertainties defined by adding the statistical and systematic sources in quadrature, are
$\mathrm{SF}^{\Pe\Pe} = 0.957 \pm 0.016$, $\mathrm{SF}^{\mu\mu} = 0.974 \pm 0.016$,
and  $\mathrm{SF}^{{\Pe}\mu} = 1.002 \pm 0.010$, for the \eepm,
\mmpm, and \empm\ \ttbar\ final states, respectively.

Systematic uncertainties arising from the lepton energy scale
 are estimated by comparing the position of the dilepton invariant-mass peak
in \dy-enriched events in data with simulations, applying only trigger and
dilepton selection criteria. The effect corresponds to a 0.3\% uncertainty on the  \ttbar\  selection efficiency for all three dilepton final states.

The uncertainties on jet energy scale (JES) and jet energy resolution (JER) affect the efficiency of jet selection.
The impact of uncertainty on JES is estimated from the change observed
in the number of selected MC \ttbar\ events after changing the jet momenta
within the JES uncertainties~\cite{JESPAS}.
Similarly, the effect of the JER uncertainty is estimated by changing the resolution on jet momenta by ${\pm}10\%$,
and then estimating the corresponding change in the number of selected MC \ttbar\ events.
The systematic uncertainty on $\met$ is discussed in Section~\ref{sec:event_selection}.%\ref{sec:metsel}.

After requiring at least one b-tagged jet, the uncertainties on the data-to-simulation factors for b tagging
 are propagated to the selection efficiency
in the simulated \ttbar\ samples. The uncertainties
on the b-tagging scale factors in \ttbar\ signal events are ${\approx}2\%$ for b jets and
${\approx}10\%$ for mistagged jets~\cite{BTV-11-004}.

The 8\% uncertainty in the inelastic proton-proton cross section at $\sqrt{s} = 7\TeV$~\cite{TOTEM73.5mb},
when propagated to the modeling of pileup, leads to an uncertainty of 0.5\%  on the yield of simulated \ttbar\ events.

Systematic uncertainties on the MC modeling of \ttbar\ production are estimated by
using \MADGRAPH with different parameter settings, such as
renormalization/factorization scales $\mu$, which are increased and decreased by a factor of 2 to estimate the effect on the calculated
\ttbar\ cross section.
Other studies change the threshold that controls the matching of partons from the matrix element with those from parton showers.
Systematic uncertainties in the \ttbar\ event selection efficiency from independent changes in the two $\mu$ scales and in the matching
of partons produce a total
uncertainty of 0.6\% for each source.
Systematic effects from the choice of parton distribution functions are also studied and
found to be negligible.

The systematic uncertainty arising from the
uncertainty on the measured values of leptonic branching fractions of the \PW\ boson
($0.1080\pm 0.0009$~\cite{PDG2012}) translates into a 1.7\% uncertainty on the number of selected MC \ttbar\ events.
This uncertainty can be reduced through a more precise measurement of the leptonic branching fractions of the \PW\,
and used to recalculate the measured cross section.

Uncertainties on the selected number of single-top-quark and diboson events
are estimated through simulation, and arise from the same kind of sources that affect the \ttbar\ signal.
The above-described detector effects
contribute to an 8\% uncertainty on the number of selected background events, which is dominated by the uncertainty on the JES.
In events with at least one b-tagged jet, the uncertainty from b tagging is ${\approx}15\%$ for diboson and 5\% for single-top-quark events, arising mainly from misidentification of light-flavor jets.
In addition, an uncertainty in the cross section for single-top-quark and diboson backgrounds, estimated to be ${\approx}20\%$~\cite{CMStopPublication4, ATLASWWPublication,
ATLASWZPublication,ATLASZZPublication}, is added in quadrature.

Table \ref{tab:all_sys_max_signal}  summarizes the impact of relative systematic uncertainties from selection efficiencies in simulated \ttbar\ signal events.
Uncertainties on background contamination from DY processes and from events with non-prompt leptons
 are estimated from data, as described in Sections~\ref{sec:dydata} and~\ref{sec:fakesdata}, and given in Table~\ref{tab:xsecChannels} of Section~\ref{sec:yields}.

\begin{table}[ht]
\begin{center}
\topcaption{Summary of the relative (\%) systematic uncertainties on the
number of signal \ttbar\ events, after applying the full selection criteria, both, before b tagging
and with at least one b-tagged jet in the event.
Combined uncertainties are listed for the sum of contributions from the three dilepton channels, except for lepton efficiencies,
which are given separately for \eepm, \mmpm, and \empm\ events.
}
\begin{tabular}{lc|c}
\hline
\hline
 & \multicolumn{2}{c} {Uncertainty on number of \ttbar\ events (\%)} \\
Source & Without b tagging  & ${\ge}1$  b-tagged jet \\
\hline
Luminosity           & 2.2 & 2.2 \\ \hline
Lepton efficiencies  & $1.7$ (ee) / $1.7$ ($\mu\mu$) / $1.0$ ($\Pe\mu$) & $1.7$ (ee) / $1.7$ ($\mu\mu$) / $1.0$ ($\Pe\mu$) \\
Lepton energy scale  & $0.3$ & $0.3$ \\
Jet energy scale  & $1.8$  & $1.9$  \\
Jet energy resolution  & $0.5$ & $0.3$  \\
\met\ efficiency & $1.4$ & $1.3$  \\
b tagging & - & $0.7$  \\
Pileup & $0.5$ & $0.5$  \\
Scale of QCD ($\mu$) & $0.6 $ & $0.6$  \\
Matching partons to showers & $0.6$ & $0.6$  \\
W branching fraction & $1.7$ &$1.7$  \\
\hline
\hline
\end{tabular}
\label{tab:all_sys_max_signal}
\end{center}
\end{table}

\section{Event yields and distributions}
\label{sec:yields}
In this section, the predicted distributions from simulation are compared with those from data.
When possible, the yields predicted for background are estimated using data, as discussed in Section~\ref{sec:bkgd}. The remaining backgrounds, as well as the
distributions in simulated \ttbar\  events, are scaled to the measured integrated  luminosity, assuming a \ttbar\ cross section of 164\unit{pb}
\cite{Langenfeld:20011}.
In the figures below, the hatched regions correspond to uncertainties on the predicted event yields.

\begin{figure}[htbp]
\begin{center}
  \includegraphics[width=0.49\textwidth]{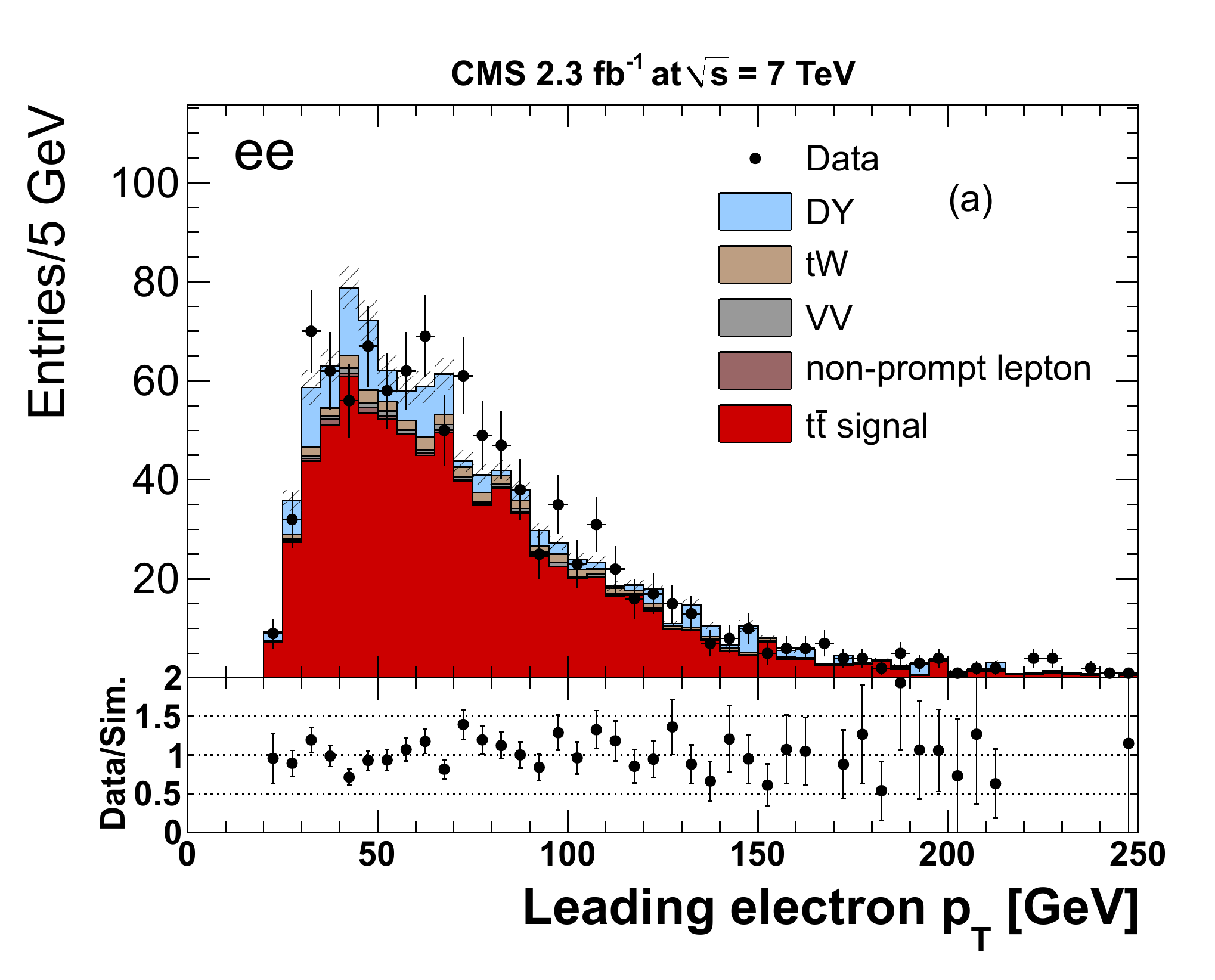}
  \includegraphics[width=0.49\textwidth]{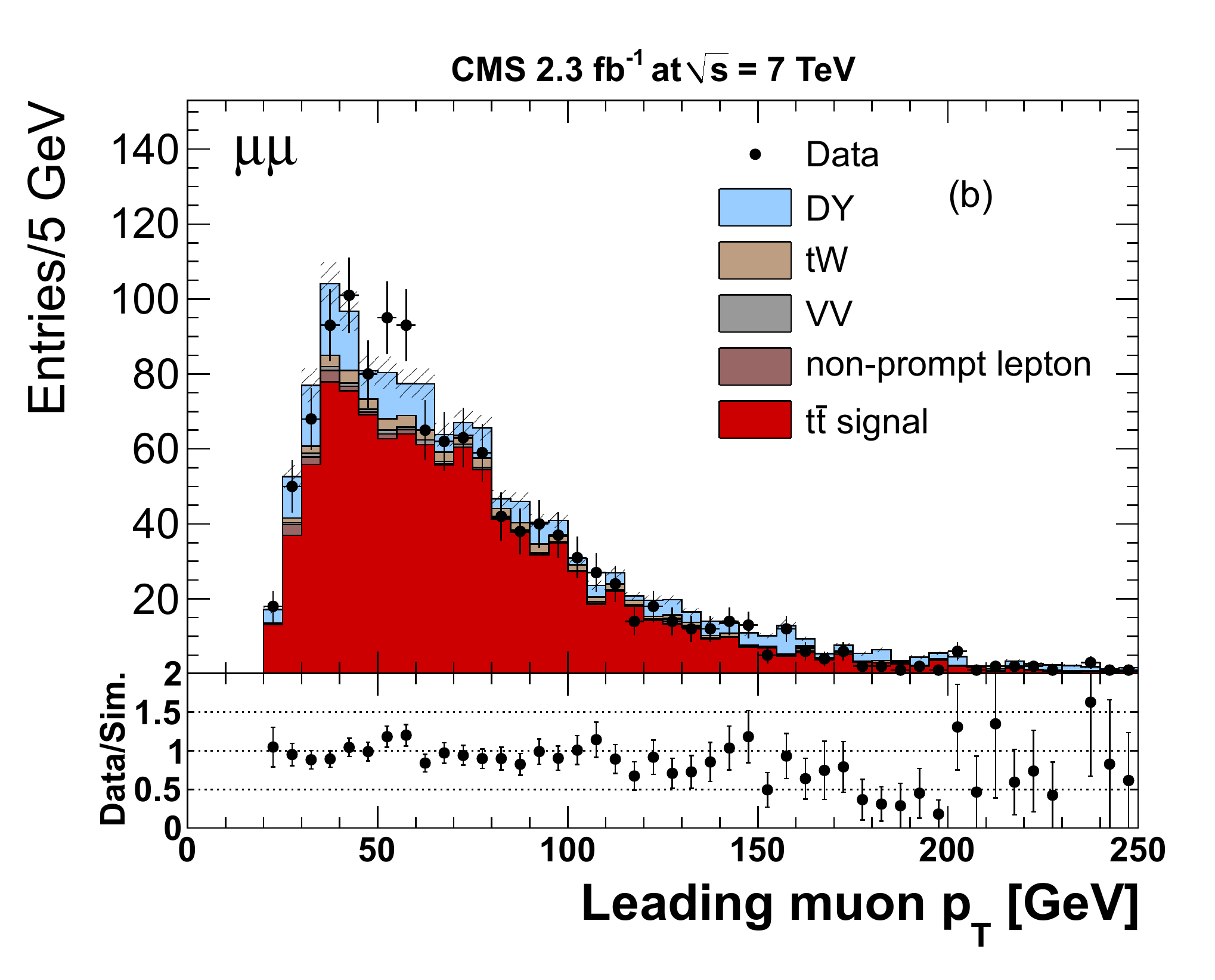} \\
  \includegraphics[width=0.49\textwidth]{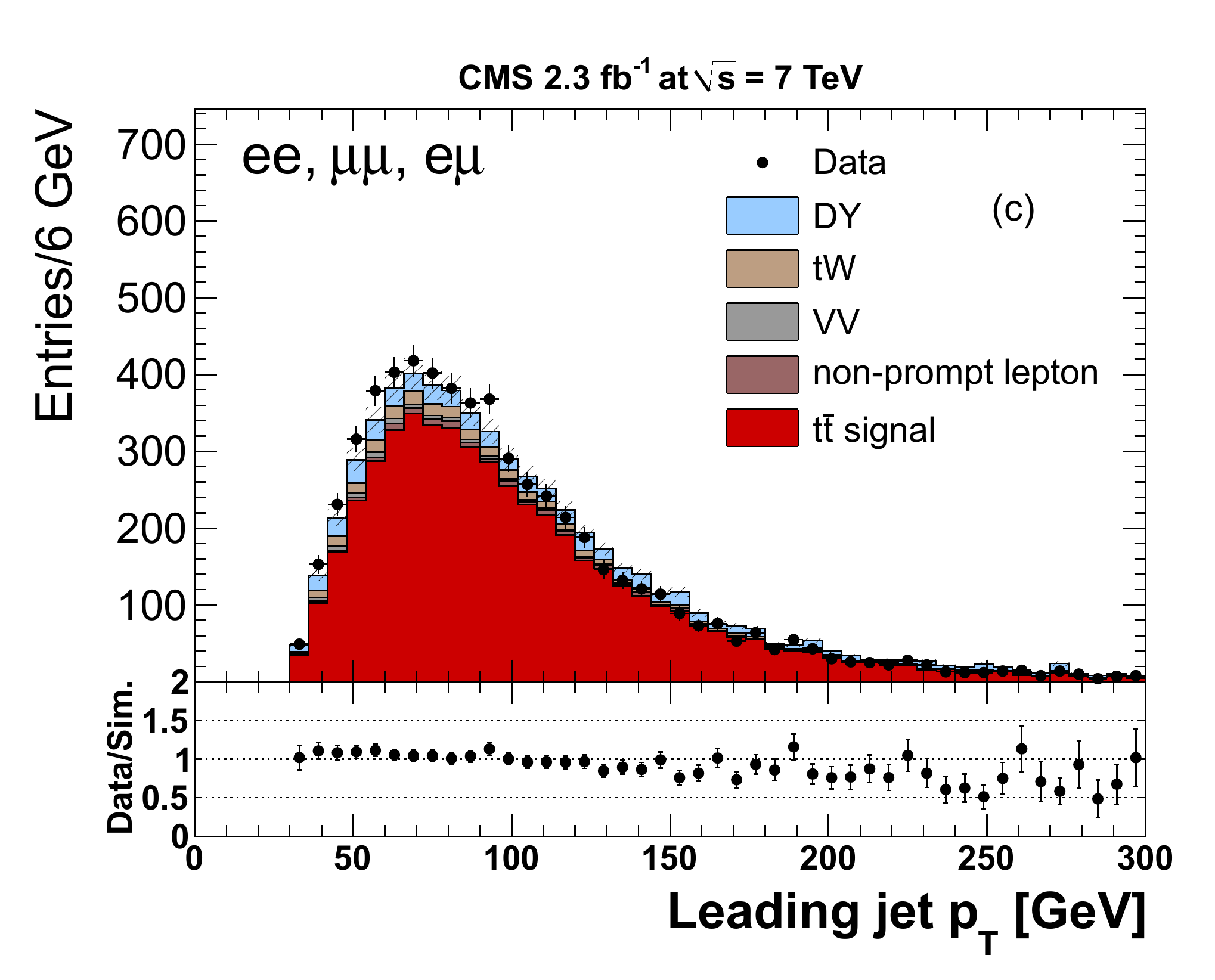} \\
\caption{The $\pt$ distributions for (a) electrons, (b) muons,  and  (c) jets of the largest-$\pt$ after the jet multiplicity selection. The expected distributions for \ttbar\ signal and individual backgrounds are shown by the histograms,
and include all data-based corrections.  A \ttbar\ cross section of 164\unit{pb} is used to normalize the simulated
\ttbar\ signal.
In this and all following figures, the hatched regions show the total
uncertainties on the sum of the \ttbar\ and background predictions.
The ratios of data to the sum of the \ttbar\ and background predictions  are
given at the bottom of each panel and the corresponding error bars include statistic and systematic uncertainties added in quadrature.}
\label{fig:pt1}
\end{center}
\end{figure}

\begin{figure}[htbp]
\begin{center}
  \includegraphics[width=0.49\textwidth]{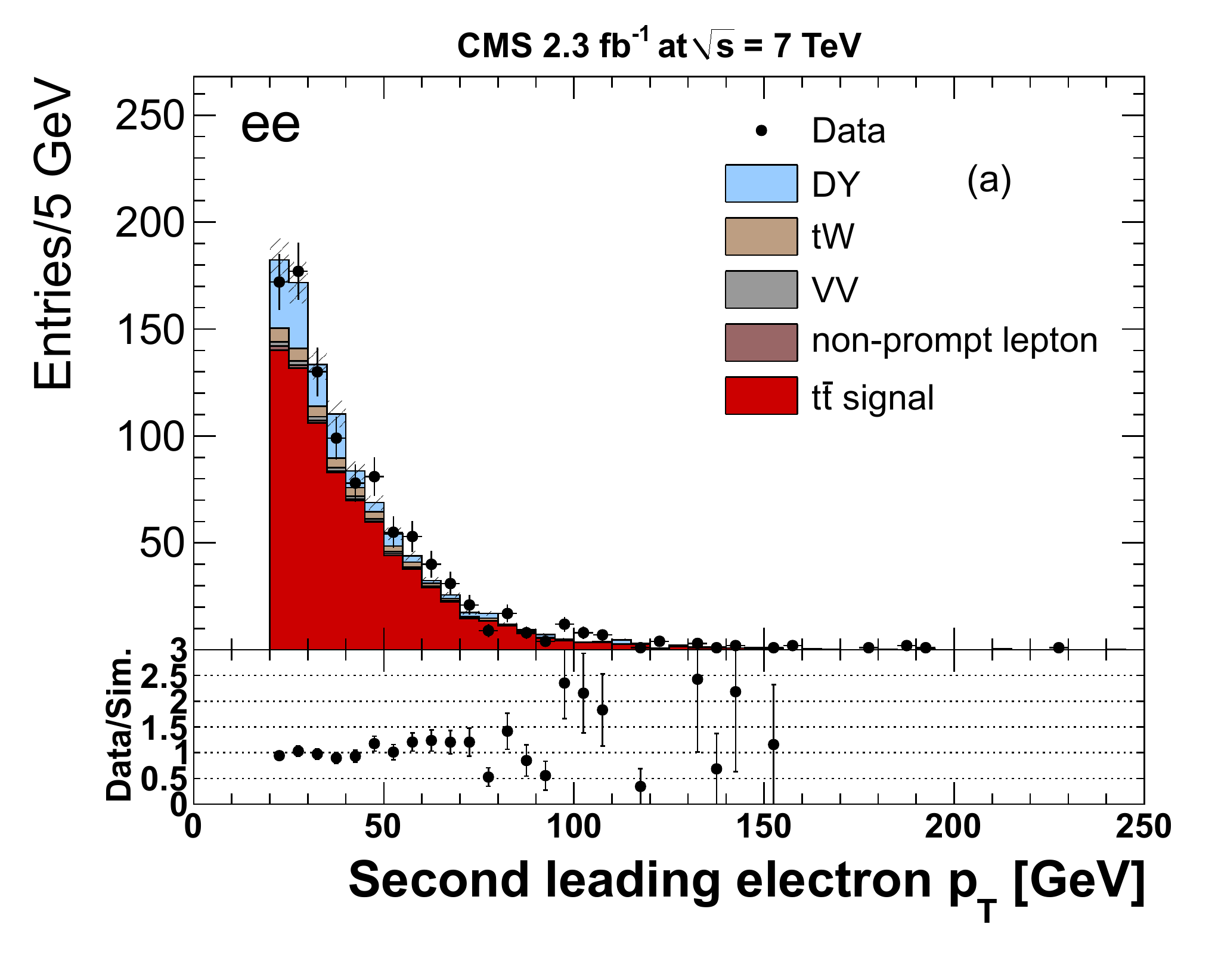}
  \includegraphics[width=0.49\textwidth]{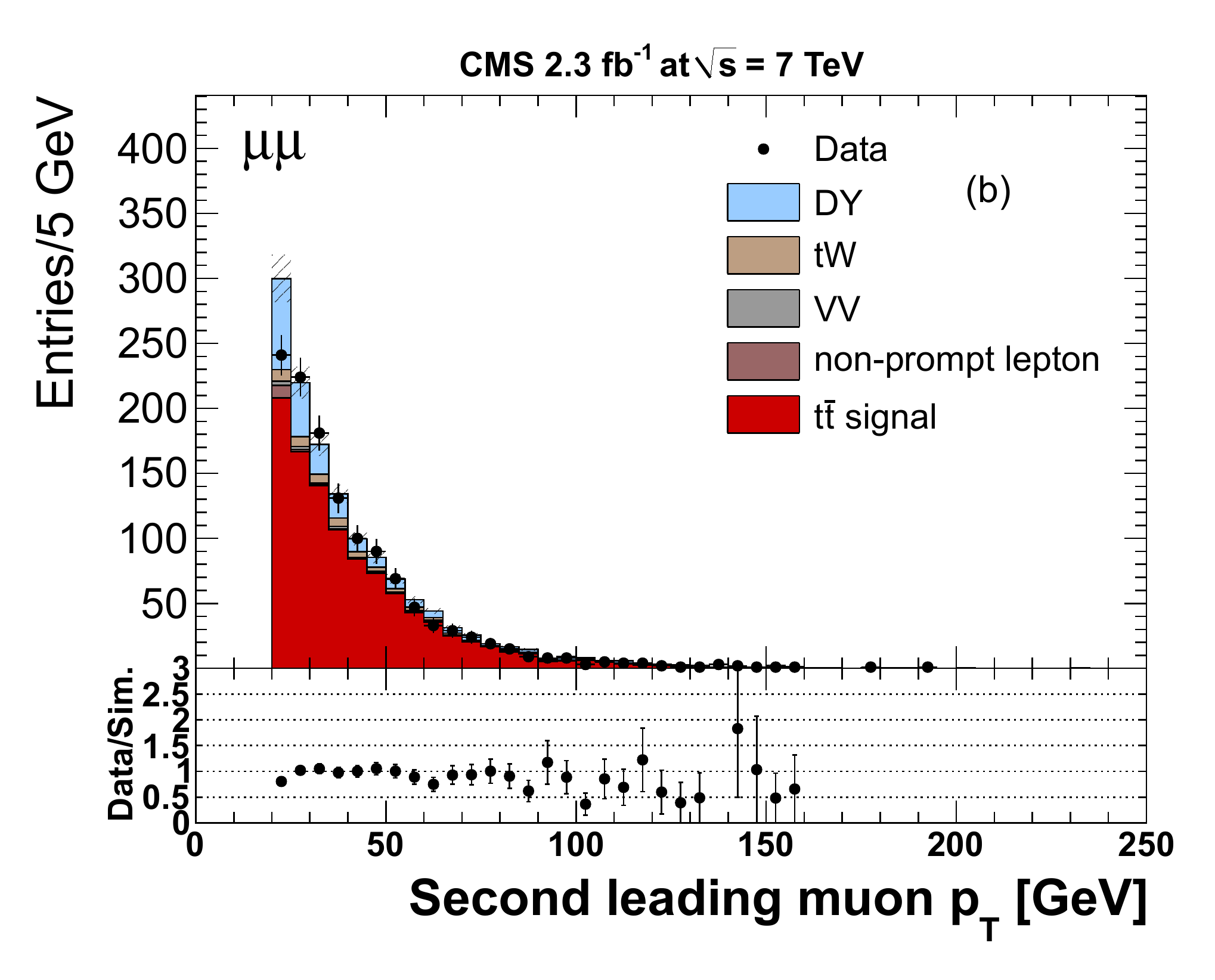} \\
  \includegraphics[width=0.49\textwidth]{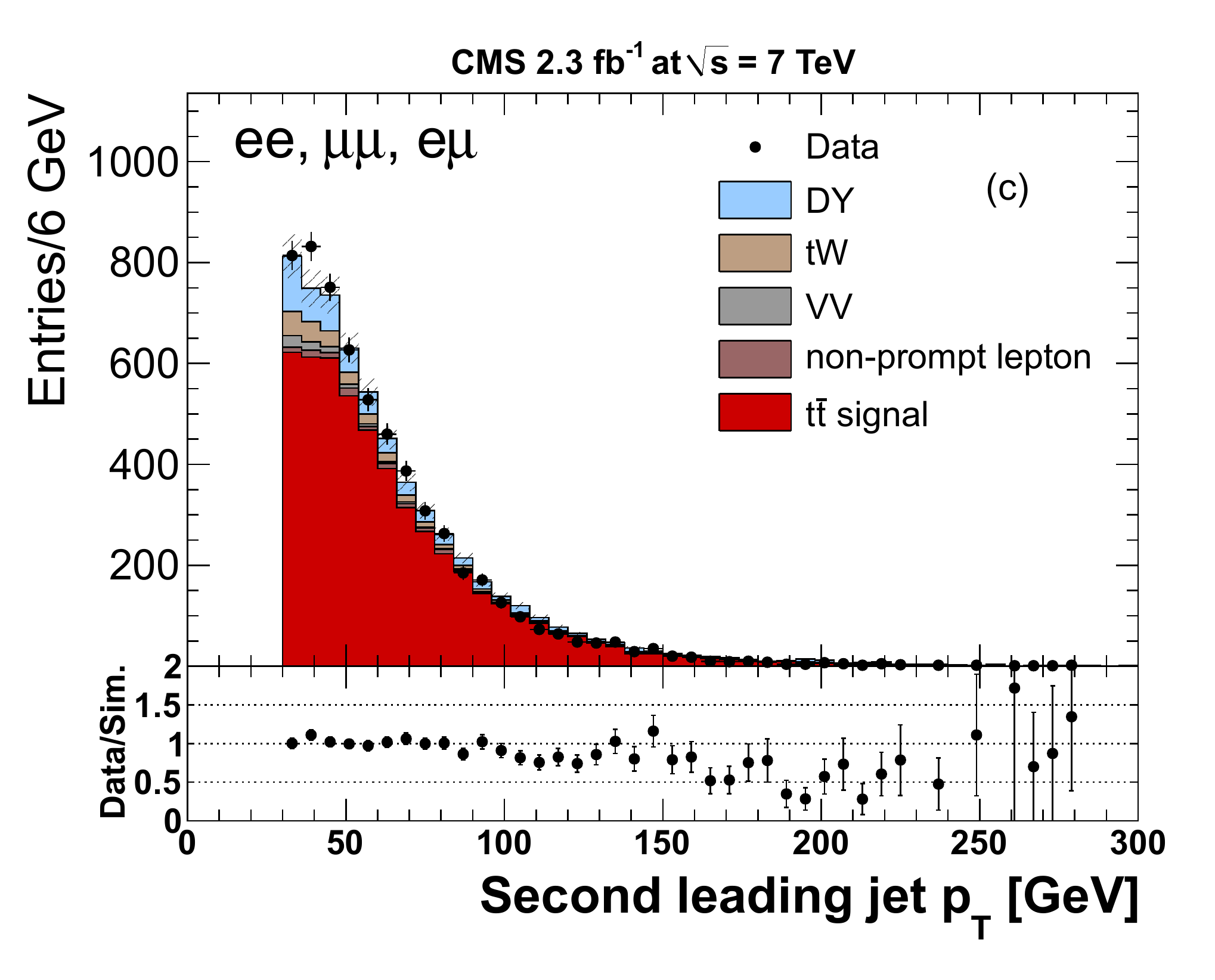} \\
\caption{Same as Fig.~\ref{fig:pt1}, but for the second-largest $\pt$
electrons, muons and jets in each event.}
\label{fig:pt2}
\end{center}
\end{figure}

\begin{figure}[htbp]
\begin{center}
  \includegraphics[width=0.49\textwidth]{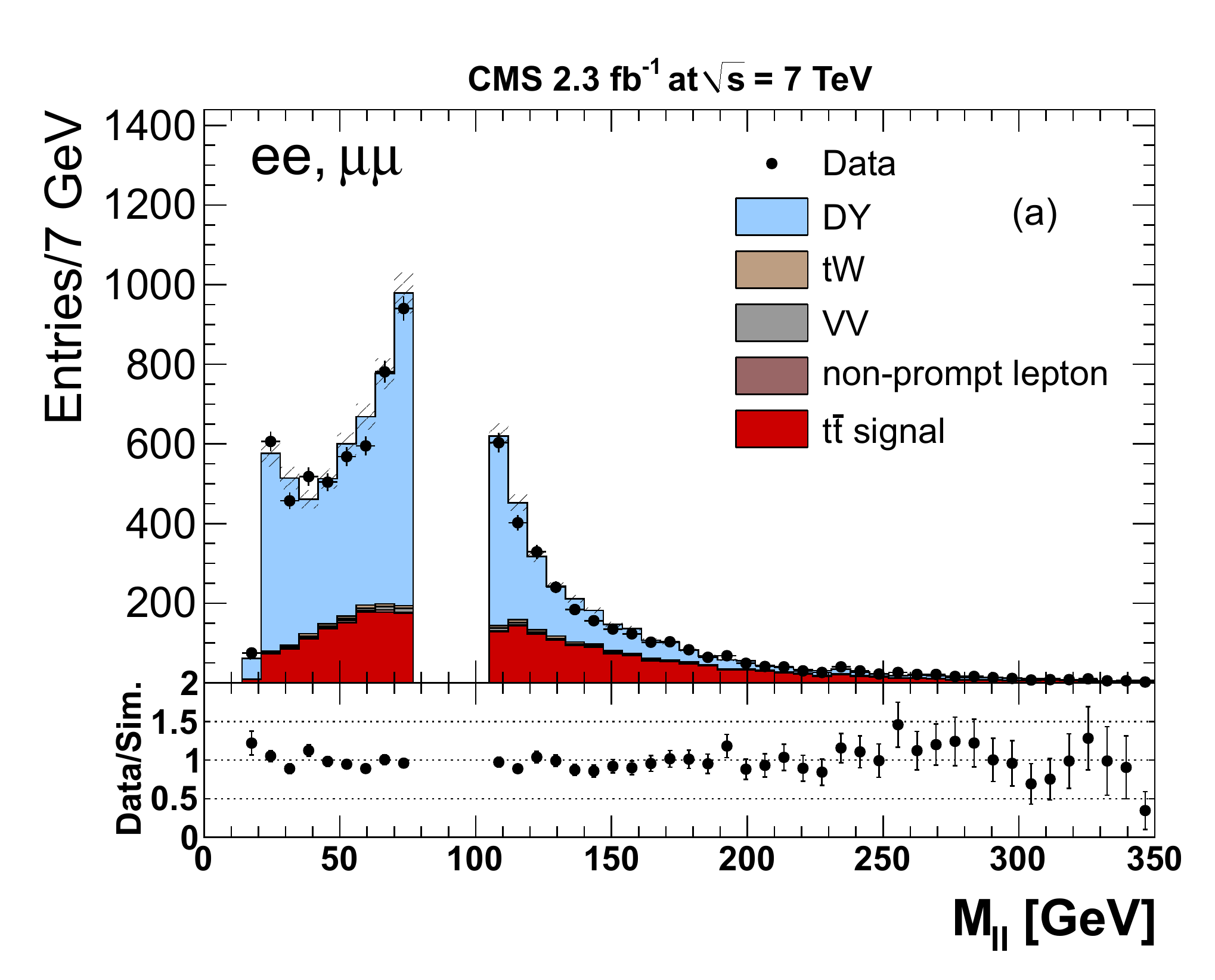}
  \includegraphics[width=0.49\textwidth]{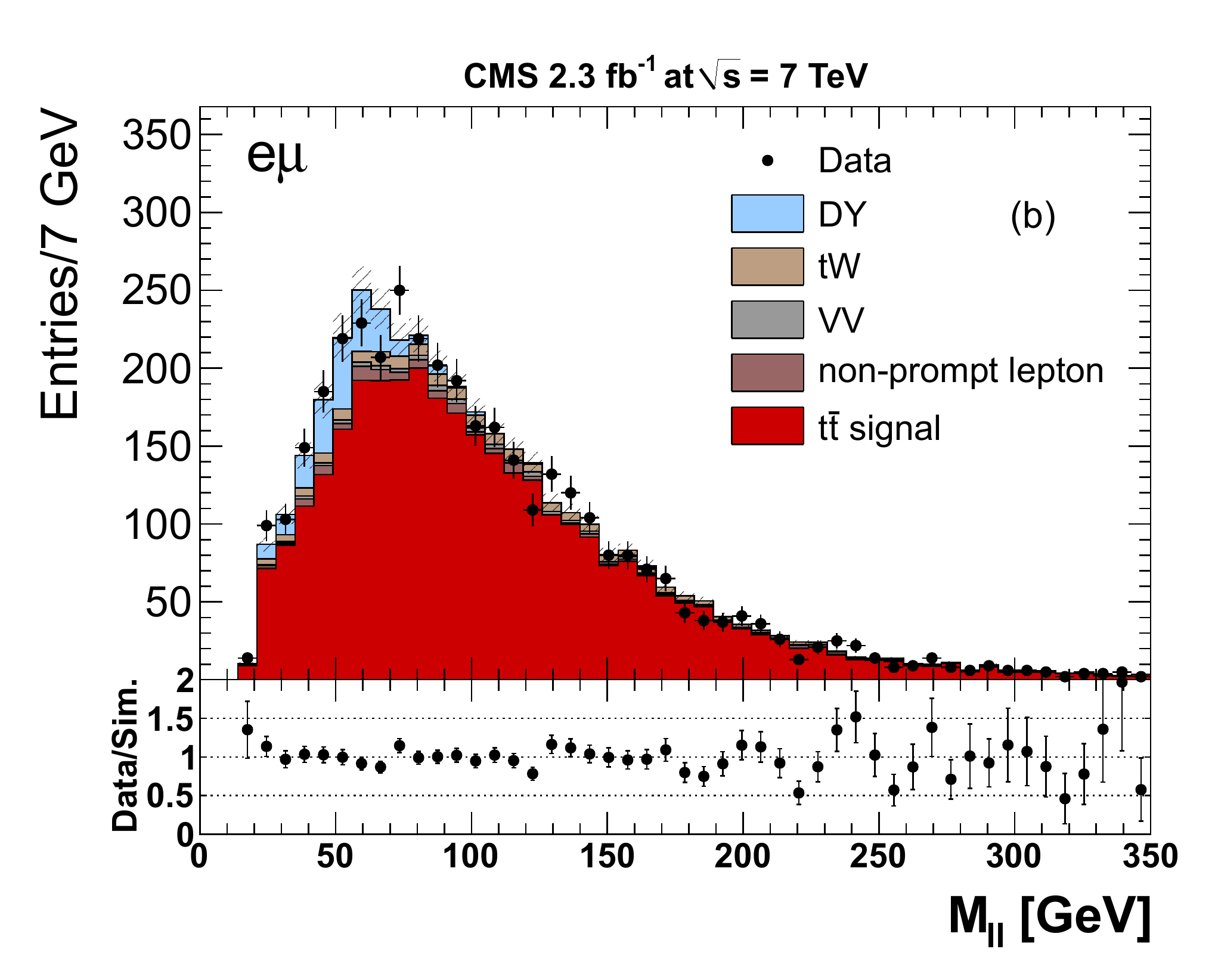} \\
\caption{Same as Fig.~\ref{fig:pt1}, but for the dilepton invariant-mass distribution of (a) the  sum of the \eepm\ and \mmpm\ channels, and
 (b) the \empm\ channel. The gap in the former distribution reflects the requirement that removes dileptons from Z decay.}
\label{fig:invmass}
\end{center}
\end{figure}

\begin{figure}[htbp]
\begin{center}
  \includegraphics[width=0.49\textwidth]{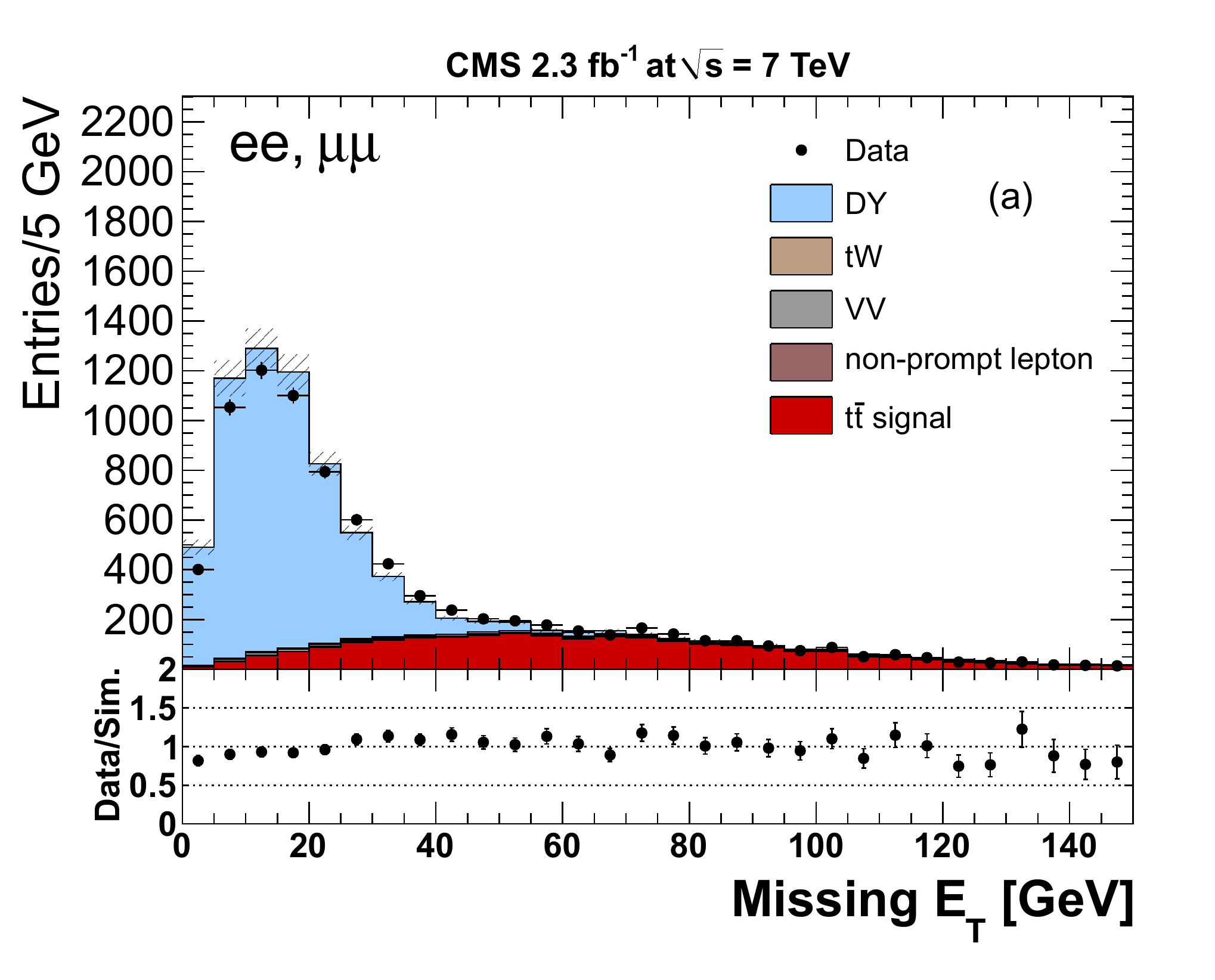}
  \includegraphics[width=0.49\textwidth]{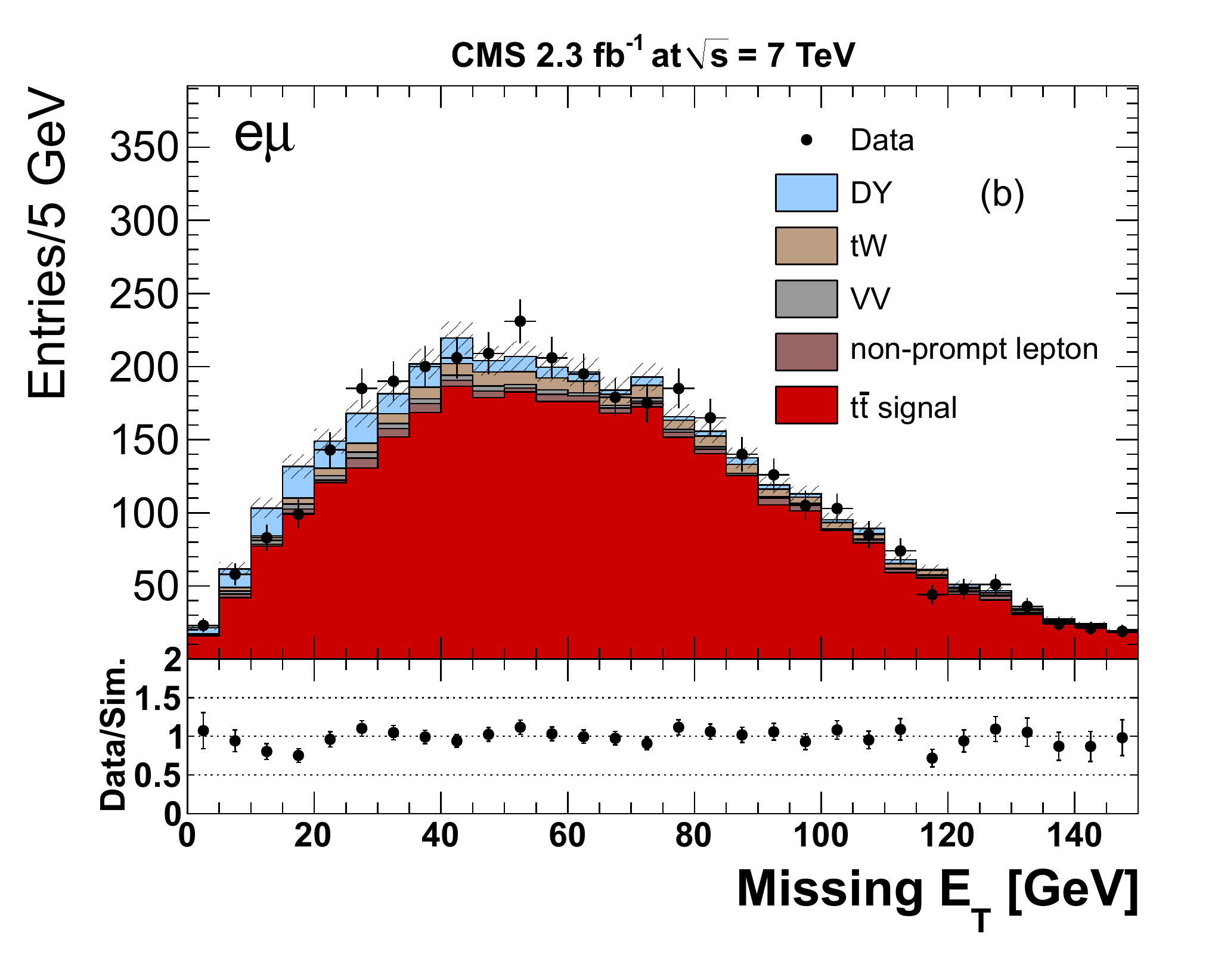} \\
\caption{The \met\ distributions after the selection on jet multiplicity. 
Details on the distributions are same as for Fig.~\ref{fig:pt1}.}
\label{fig:met}
\end{center}
\end{figure}

\begin{figure}[htbp]
\begin{center}
  \includegraphics[width=0.49\textwidth]{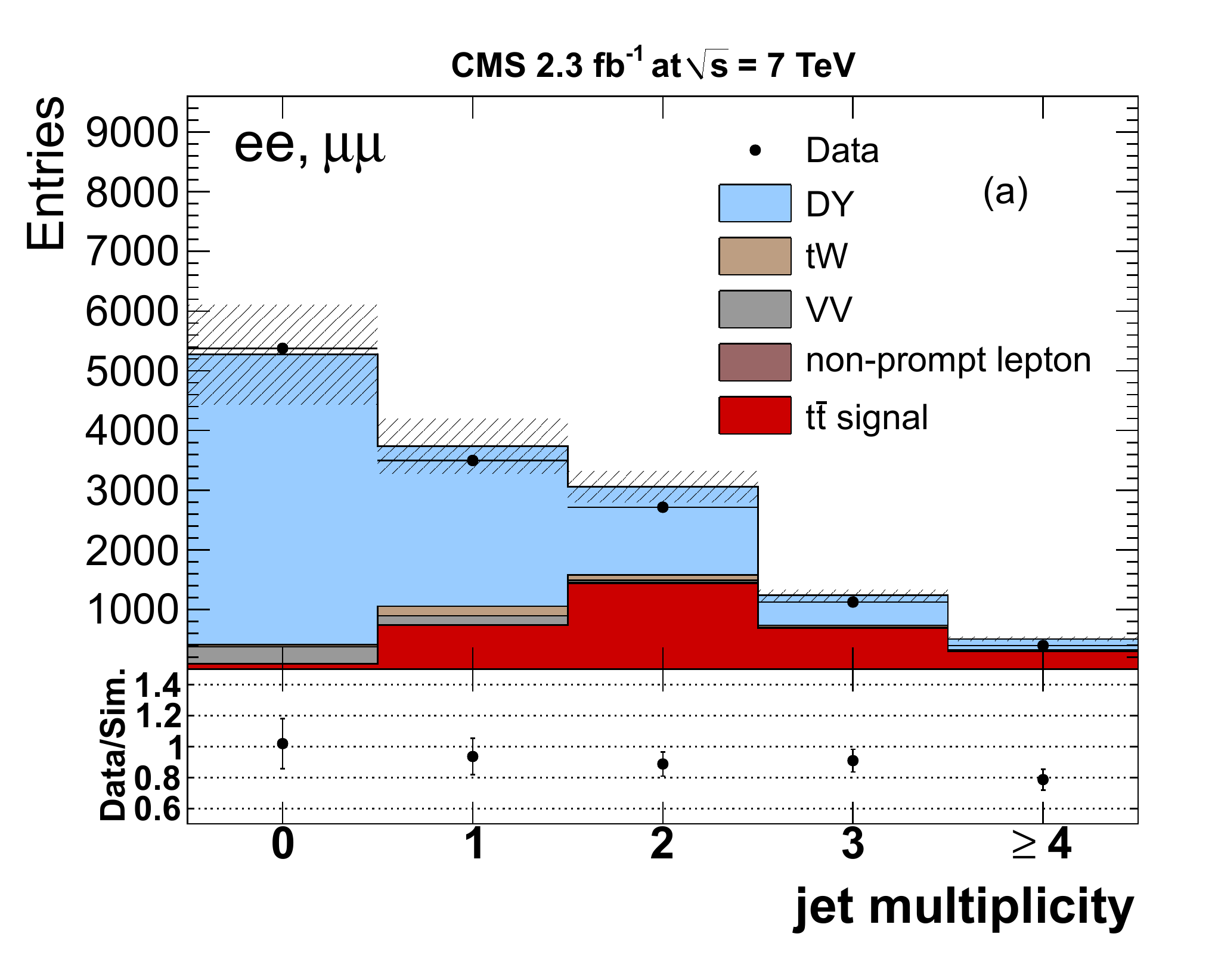}
  \includegraphics[width=0.49\textwidth]{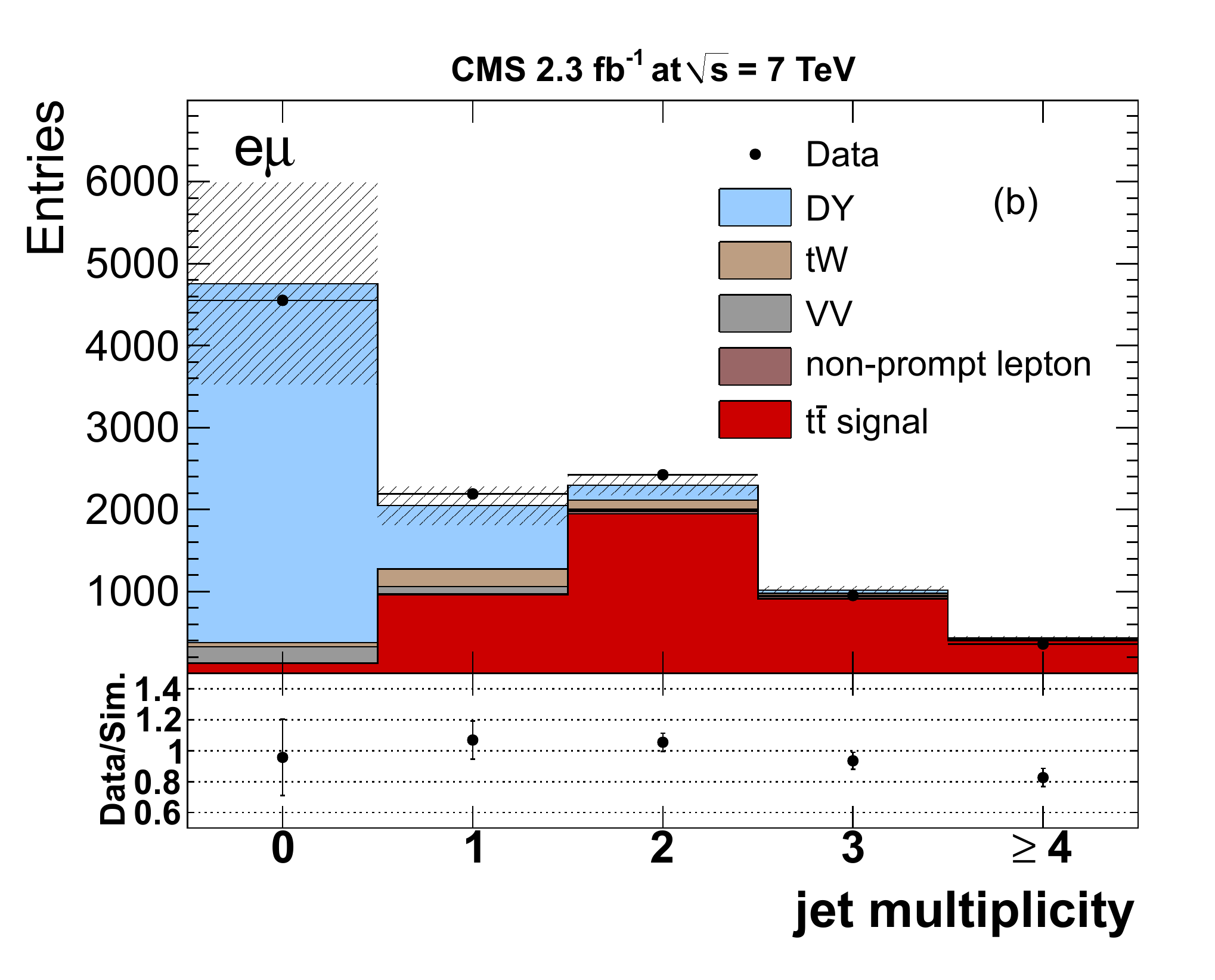}\\
\caption{The jet multiplicity for events passing the dilepton and \met\
criteria, but  before the b-tagging requirement,
 for (a) the sum of \eepm\ and  \mmpm channels, and (b) the \empm\ channel.}
\label{fig:njetsDD}
\end{center}
\end{figure}

\begin{figure}[htbp]
\begin{center}
  \includegraphics[width=0.49\textwidth]{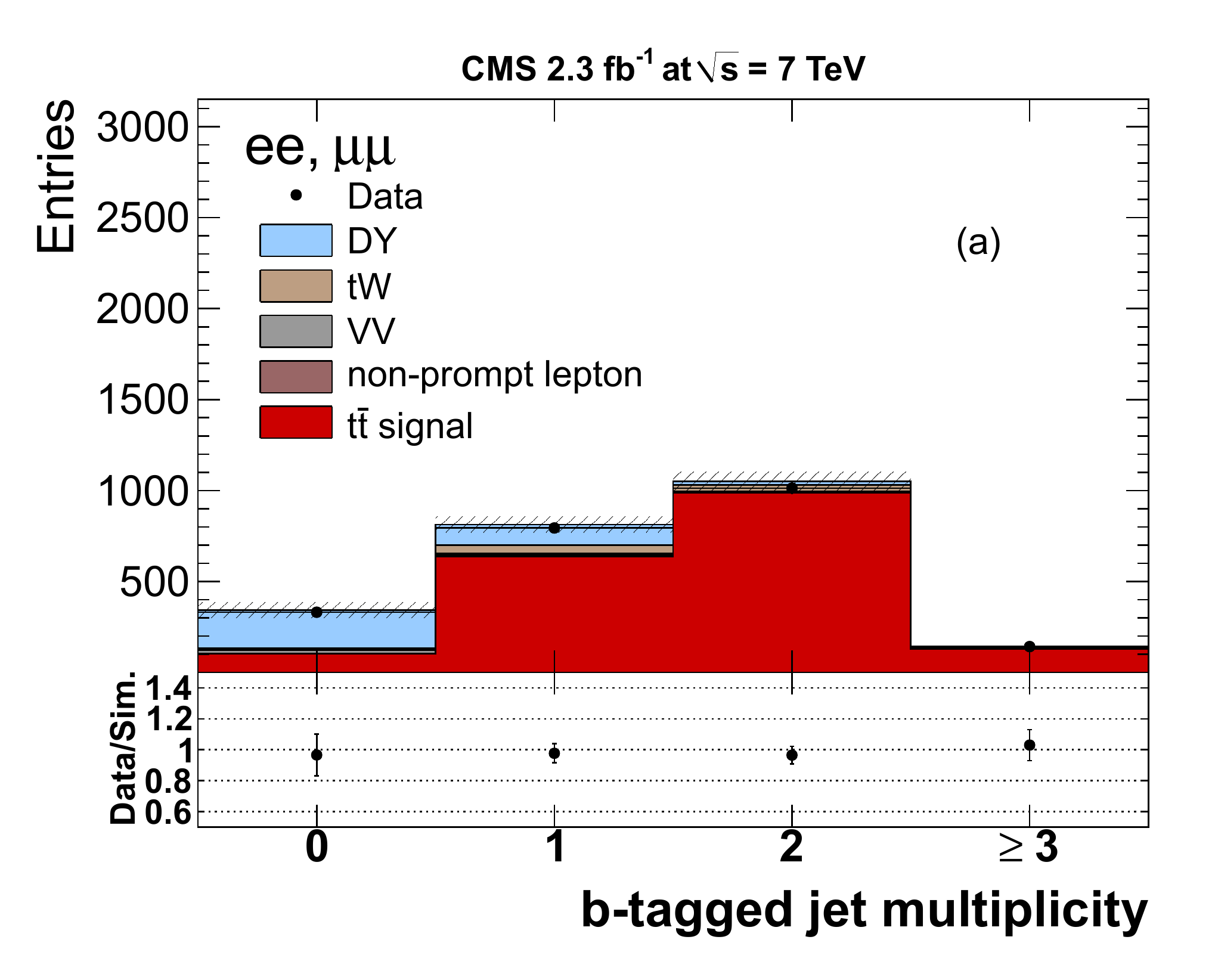}
  \includegraphics[width=0.49\textwidth]{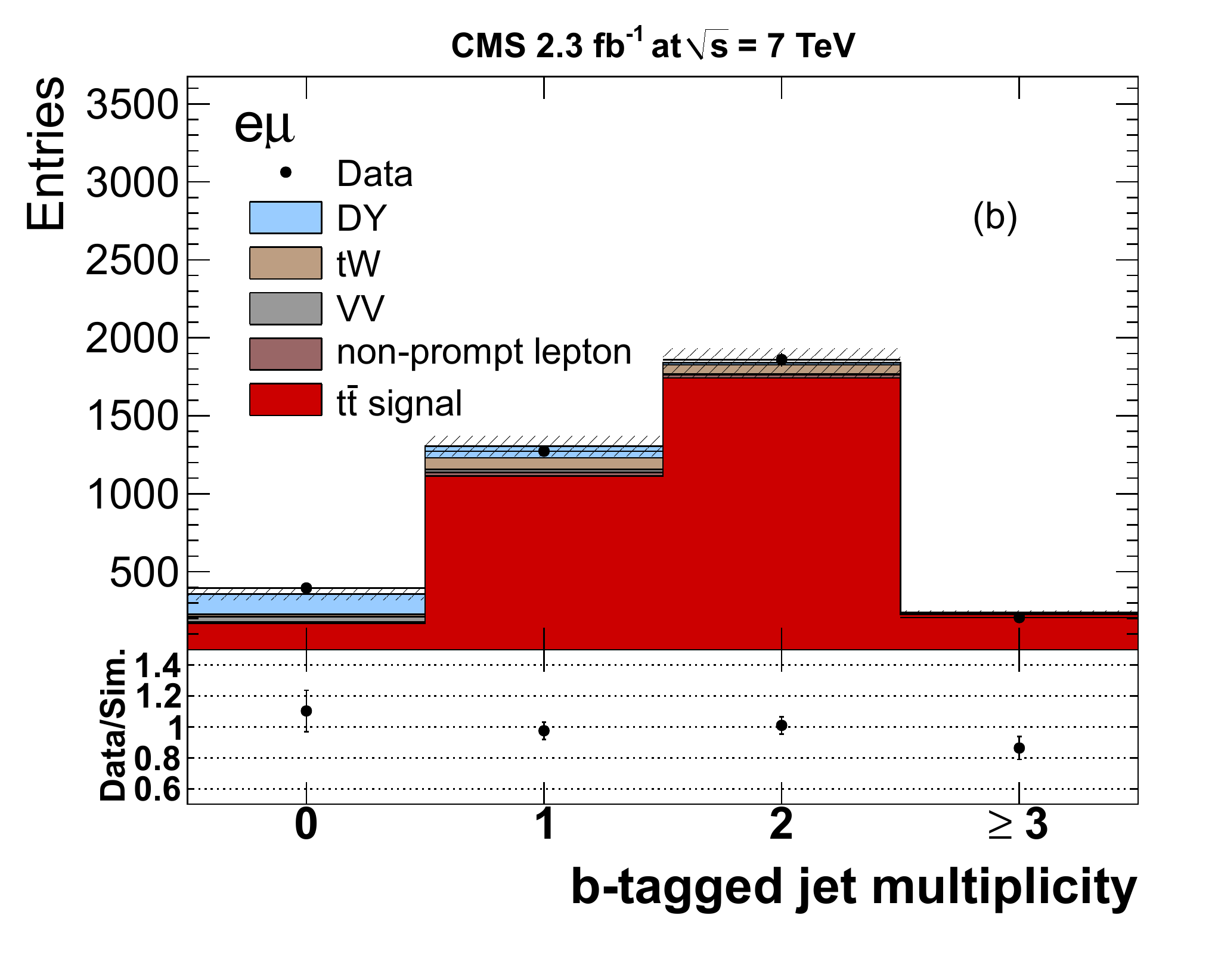} \\
\caption{The multiplicity of b-tagged jets in events passing
full event selections
 for (a) the summed \eepm\ and \mmpm\ channels, and (b) the \empm\  channels.}
\label{fig:nbtags}
\end{center}
\end{figure}

Transverse momentum distributions of leptons and jets are shown, respectively, for the largest (leading) $\pt$ and next-to-largest (next-to-leading) $\pt$
 of each set of these objects, after jet multiplicity selections, separately for the two $\ell^+ \ell^-$ and for all combined dilepton channels,
 in Figs.~\ref{fig:pt1} and~\ref{fig:pt2}, respectively.
The ratio of the data to the sum of simulations for the signal and backgrounds are shown in the bottom panels of each of the figures.
The corresponding \mll\ and \met\ distributions and their ratios to expectations are shown, respectively, in Figs.~\ref{fig:invmass} and~\ref{fig:met},
separately for the combined two $\ell^+ \ell^-$ channels and for the \empm\ channel.
The multiplicity of selected jets is presented in Fig.~\ref{fig:njetsDD},  after applying the dilepton and \met\ selections.
The observed number of events with less than two jets is used
to check the reliability of background predictions.
The multiplicity of jets that pass the b-tagging selections is given in Fig.~\ref{fig:nbtags},
again separately for the summed  \eepm\ and \mmpm\ channels, and for the \empm\ channel.
The observed number of events and the expectation from simulations and data-based predictions
 are presented in Table~\ref{tab:xsecChannels}, for a \ttbar\ cross section of 164\unit{pb}.
The uncertainties shown account for statistical and systematic sources. The predicted and observed number of events are consistent within their
uncertainties. Also shown in Table~\ref{tab:xsecChannels} are the acceptances for simulated \ttbar\ signal events, including the two-lepton branching fractions, with and
without requiring at least one b-tagged jet.

\begin{table}[htb]
\begin{center}
\topcaption{Number of dilepton events in the \eepm, \mmpm,  and \empm channels
after applying the event-selection criteria: (a)
without requiring a b-tagged jet, and (b) requiring  at least one b-tagged
jet. The results are given for the individual sources of
 background, \ttbar\ signal for $\sigma_{\rm t\bar{t}}$ = 164\unit{pb}, and the data.
The uncertainties reflect statistical and systematic uncertainties added in quadrature.
Panel (c) gives the \ttbar\ acceptance multiplied by the selection efficiency and by the branching fractions $B$ (in \%) of \ttbar\ to two-lepton states, estimated using \ttbar\ simulated events.
}
\begin{tabular}{lr@{}c@{}lr@{}c@{}lr@{}c@{}l}
\hline
\hline
         & \multicolumn{9}{c}{(a) Number of events} \\
         &\multicolumn{9}{c}{Without b-tagging selection} \\
Source   &  \multicolumn{3}{c} { \eepm\ }        &  \multicolumn{3}{c} { \mmpm\ } & \multicolumn{3}{c} { \empm\ } \\ \hline
Drell-Yan        & 136 & $\pm$ & 29
                 & 217 & $\pm$ & 45
                 & 220 & $\pm$ & 46  \\
Nonprompt leptons   & 9  & $\pm$ & 6
                    & 15 & $\pm$ & 7
                    & 85 & $\pm$ & 20  \\
Diboson          & 14 & $\pm$ & 4
                 & 16 & $\pm$ & 4
                 & 55 & $\pm$ & 13  \\
Single top       & 42 & $\pm$ & 9
                 & 53 & $\pm$ & 11
                 & 156 & $\pm$ & 32 \\
\hline
Total background               & 201 & $\pm$ & 31
                               & 301 & $\pm$ & 47
                               & 516 & $\pm$ & 61 \\
\ttbar signal   & 801  & $\pm$ & 34
                & 1041 & $\pm$ & 43
                & 3253 & $\pm$ & 126 \\
Total predicted	& 1002 & $\pm$ & 46
                & 1342 & $\pm$ & 64
		& 3769 & $\pm$ & 140\\
\hline
Data  &  \multicolumn{3}{c}{1021}  &  \multicolumn{3}{c}{1259}  &  \multicolumn{3}{c}{3734}  \\ \hline
\hline
 & \multicolumn{9}{c}{(b) $\ge1$b-tagged jet} \\
Source   &  \multicolumn{3}{c} { \mmpm\ }        &  \multicolumn{3}{c} { \eepm\ } & \multicolumn{3}{c} { \empm\ } \\ \hline
Drell-Yan        & 62 & $\pm$ & 16
                 & 82 & $\pm$ & 21
                 & 89 & $\pm$ & 19 \\
Nonprompt leptons             & 2.4 & $\pm$ & 4.8
                              & 10.0 & $\pm$ & 5.5
                              & 50 & $\pm$ & 15 \\
Diboson          & 5.7 & $\pm$ & 1.4
                 & 6.1 & $\pm$ & 1.5
                 & 22.3 & $\pm$ & 5.3 \\
Single top       & 37.5 & $\pm$ & 7.8
                 & 47.0 & $\pm$ & 9.8
                 & 140 & $\pm$ & 29 \\
\hline
Total background & 108 & $\pm$ & 18
                 & 145 & $\pm$ & 24
                 & 301 & $\pm$ & 38 \\
\ttbar signal   & 759  & $\pm$ & 33
                & 991  & $\pm$ & 42
                & 3082 & $\pm$ & 122 \\
Total predicted &  867  & $\pm$ & 38
                & 1136 & $\pm$ & 48
		& 3383 & $\pm$ & 128\\
\hline
Data  &  \multicolumn{3}{c}{875}  &  \multicolumn{3}{c}{1074}  &  \multicolumn{3}{c}{3339}  \\ \hline
\hline
         &\multicolumn{9}{c}{(c) \ttbar\ acceptance $\times$ eff.$\times$ $B$ (\%)} \\
b-tagging selection   &  \multicolumn{3}{c} { \eepm\ }        &  \multicolumn{3}{c} { \mmpm\ } & \multicolumn{3}{c} { \empm\ } \\ \hline
\hline
No selection       & 0.22 & $\pm$ & 0.01 & 0.28 & $\pm$ & 0.01 & 0.87 & $\pm$ & 0.04 \\ \hline
$\ge1$b-tagged jet & 0.20 & $\pm$ & 0.01 & 0.27 & $\pm$ & 0.01 & 0.83 & $\pm$ & 0.04 \\ \hline
\hline
\end{tabular}
\label{tab:xsecChannels}
\end{center}
\end{table}

\section{Measuring the \texorpdfstring{\ttbar}{t t-bar} production cross section}
\label{sec:measure}

The \ttbar\ production cross section
is measured using a profile likelihood-ratio method described in Section~\ref{sec:results_channels_plr}, while
Section~\ref{sec:results_channels_counting} discusses a counting analysis used as a cross-check. The latter is based on
counting the number of events in the data  that survive an additional selection, which requires at least one b-tagged jet, and is
referred to as a counting analysis in what follows.
For ease of comparison with previous measurements, the value of ${m_\cPqt}$= 172.5\GeV is used
to extract the \ttbar\ cross section. The dependence of $\sigma_{\ttbar}$ on ${m_\cPqt}$ is discussed in Section~\ref{sec:summary}.

\subsection{Cross-section measurement using a profile likelihood ratio}
\label{sec:results_channels_plr}

A profile likelihood ratio \cite{plr1, plr2} is used to measure
 $\sigma_{\ttbar}$ in the individual \eepm, \mmpm, and \empm\ dilepton channels, as well as in their combination.
The minimum value of $-2 \ln[\mathcal{R}(\sigma_{\ttbar})]$ is extracted from a scan over a wide range of $\sigma_{\ttbar}$,
where $\mathcal{R}$ is defined as:
\begin{equation}
\mathcal{R}(\sigma_{\ttbar}) = \frac{L( \sigma_{\ttbar}, \{ \hat{\hat{U_i}} \}) }
    { L(\hat{ \sigma}_{\ttbar}, \{ \hat{U_i} \})} ,
\label{PLRrefeq}
\end{equation}
with $L$ being the  likelihood
 function with one free parameter, $\sigma_{\ttbar}$, and a set of other parameters $\{U_i\}$, called ``nuisance parameters'', that describe
the estimated systematic uncertainties in the measurement.
The set $\hat{\hat{\{U_i\}}}$ denotes the conditional maximum-likelihood (ML) estimates of the $\{U_i\}$
that depend on the specific value of $\sigma_{\ttbar}$.
The denominator corresponds to the maximized unconditional likelihood function,
with its ML estimators defined as $\hat{\sigma}_{\ttbar}$ and $\hat{\{U_i\}}$.
The presence of the nuisance parameters broadens the distribution of the  $\mathcal{R}$ function, which
reflects the loss of information on $\sigma_{\ttbar}$ from the presence of systematic uncertainties.

\begin{figure}[hbtp]
  \begin{center}
       \includegraphics[width=0.65\textwidth]{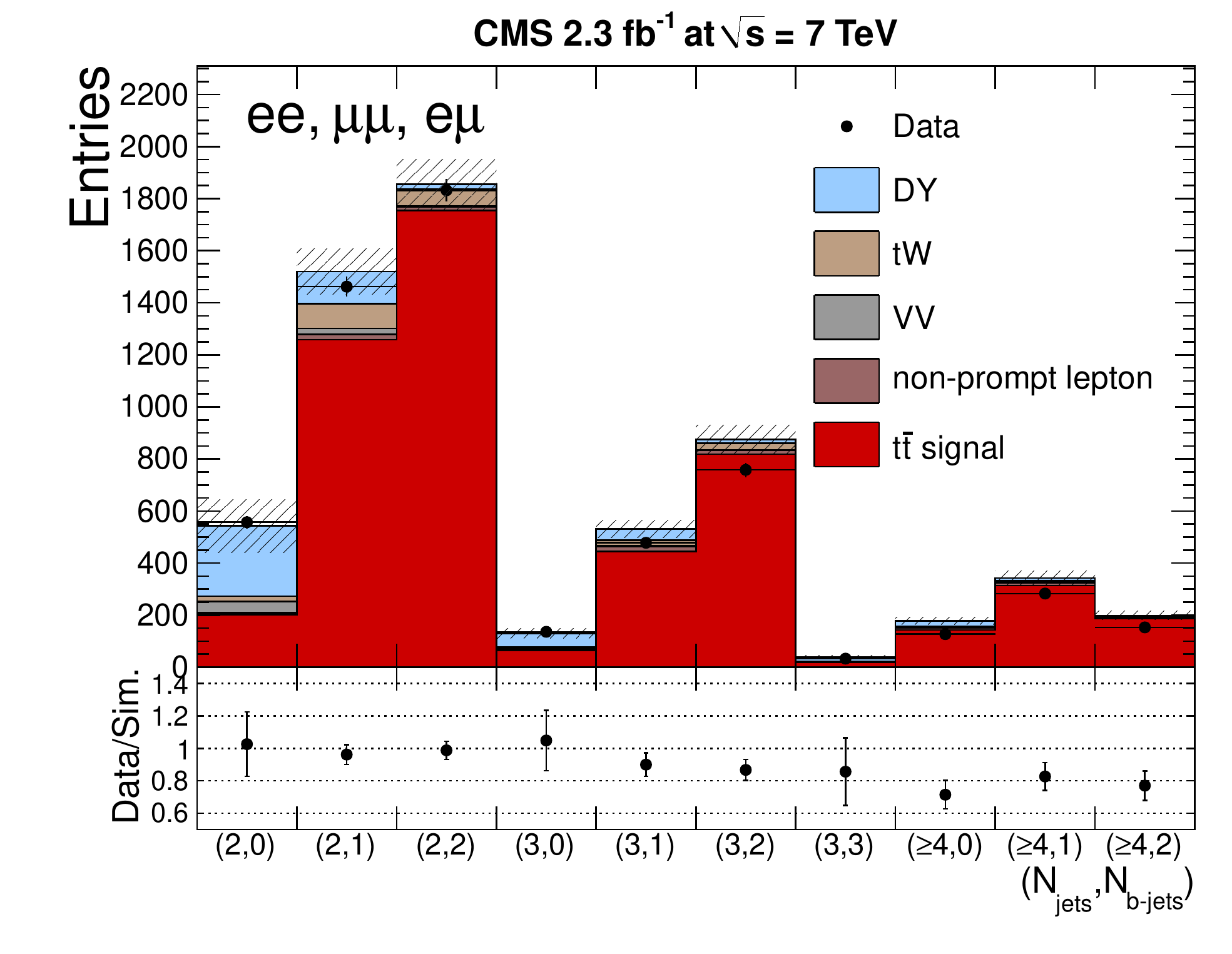}
  \end{center}
    \caption{
    Number of events selected for the  three combined dilepton channels, as a function of the number of jets and b-tagged jets
    ($N_\text{jets}$,$N_\text{\cPqb-jets}$) in each event.
    The data are shown by the dots, while the predicted \ttbar\ and the contributing backgrounds are shown by the histograms. The hatched area
    corresponds to statistical and systematic uncertainties on the \ttbar\ and on the background predictions taken in quadrature.
    The ratios of data to the sum of the \ttbar\ and background predictions are given at the bottom, with the error bars
    again giving the statistical and
    systematic uncertainties taken in quadrature.
    }
    \label{fig:njetsnbjets}
\end{figure}

We consider a  likelihood defined by a
probability density function  binned in a 2-dimensional (2D) space of  jet multiplicity
($N_\text{jets}$) and multiplicity of b-tagged jets ($N_\text{\cPqb-jets}$).
The predicted 2D distributions (or ``templates'') are obtained by applying the event selection criteria to simulated events for each contributing process,
without imposing b-tagging requirements. The corresponding number of selected events in simulation, compared with the number selected in data,
 are shown in Fig.~\ref{fig:njetsnbjets} for different combinations of ($N_\text{jets}$,$N_\text{\cPqb-jets}$).

Without systematic uncertainties, the likelihood function represents the product of Poisson distributions that reflect the
statistical content of each bin.
To incorporate systematic uncertainties,
Gaussian terms are introduced for the nuisance parameters into  the likelihood function as constraints on $\{U_i\}$.
A systematic uncertainty
 can affect the 2D distribution by changing its normalization and its differential dependence. The effect is estimated by
means of
an interpolation between the original 2D template and
 the ones obtained by changing each nuisance parameter $U_{i}$ by $\pm$1 standard deviation.
For uncertainties on data-to-simulation SF in tagging of b jets and light-flavor jets,
the change in template for a given process is computed analytically for each bin of the distribution,
according to the combinatorial probability of tagging $N_\text{\cPqb-jets}$ jets in an event with $N_\text{jets}$ jets.

The $\mathcal{R}$ function given in Eq.~(\ref{PLRrefeq}) is used to extract the uncertainty on $\sigma_{\ttbar}$ from
a change by $\pm$0.5 of the fitted $-2\ln[\mathcal{R}]$ value relative to its minimum.
This represents the Gaussian equivalent of 68\% confidence for the combined statistical and systematic uncertainties.
The minimization of $ \ln[\mathcal{R}]$  corresponds to an
integration over the parametrized systematic uncertainties ($U_{i}$), which also reduces the overall uncertainty on $\sigma_{\ttbar}$.

The technique described for a given decay channel can be extended to the combined channels.
The \eepm, \mmpm, and \empm\ channels  are statistically independent because they correspond to independent data samples. The
likelihood function for the combined channels is expressed therefore as the product of the individual
likelihood functions.
Correlations among the channels are introduced through the nuisance parameters, which are taken to be the same for  the \eepm, \mmpm, and \empm\ channels.
The systematic uncertainties included in the PLR analysis are those described in  Section~\ref{sec:systematics} and summarized in Tables~\ref{tab:all_sys_max_signal} and~\ref{tab:xsecChannels}.

The results  for each dilepton channel and for the combined sample are presented in Table~\ref{tab:xsec}.
The uncertainties reflect statistical and systematic sources, and the uncertainty on integrated luminosity.
The measurements from the three dilepton channels are consistent within their uncertainties.
The combined result from
 the PLR method is  $161.9 \pm 2.5\,\text{(stat.)}\,{}^{+5.1}_{-5.0}\,\text{(syst.)}\pm 3.6\,\text{(lumi.)}\unit{pb}$.

\begin{table}[htb]
\begin{center}
\topcaption{Measured $\sigma_{\ttbar}$ in pb for a top quark mass of 172.5\GeV
for each of the dilepton channels, as well as for their combination. The quoted uncertainties
are, respectively, from statistical and systematic sources and the contributions from
integrated luminosity.}
\renewcommand*\arraystretch{1.3}
\begin{tabular}{lcc}
\hline
\hline
Channel & PLR method  & Counting analysis  \\
\hline
${\Pe\Pe}$           & $168.0 \pm 6.6 ^{+7.6}_{-7.0} \pm 3.7$ & 165.9 $\pm$ 6.4 $\pm$ 7.0 $\pm$ 3.6 \\
$\mu\mu$             & $156.3 \pm 5.6 ^{+7.7}_{-6.6} \pm 3.5$ & 153.8 $\pm$ 5.4 $\pm$ 6.6 $\pm$ 3.4 \\
$\Pe\mu$         & $161.9 \pm 3.1 ^{+5.8}_{-5.4} \pm 3.6$ & 161.6 $\pm$ 3.1 $\pm$ 5.6 $\pm$ 3.6 \\
\hline
Combined & $161.9 \pm 2.5 ^{+5.1}_{-5.0} \pm 3.6$ & 161.0 $\pm$ 2.6 $\pm$ 5.6 $\pm$ 3.6 \\
\hline
\hline
\end{tabular}
\label{tab:xsec}
\end{center}
\end{table}

\subsection{Measurement of \texorpdfstring{$\sigma_{\ttbar}$}{the t t-bar cross section} using a counting analysis}
\label{sec:results_channels_counting}
The \ttbar\ production cross section is also measured by counting events and applying the expression:
\begin{equation}
\sigma_{\ttbar} = \frac{N - N_B}{\mathcal{A}  {\mathcal{L}}},
\label{eqn:xsecA}
\end{equation}
where $N$ is the  total number of dilepton events observed in data, $N_B$ is the number of estimated background events,
$\mathcal{A}$ is the mean acceptance multiplied by the selection efficiency and by the branching fraction $B(\ttbar\rightarrow\text{dilepton final state})$, for all produced \ttbar\ events,
and $\mathcal{L}$ is the integrated luminosity.
The measurement is performed separately for each dilepton channel and the individual channels are combined using a BLUE procedure ~\cite{BLUE}.
While the PLR method adjusts the jet and the b-tagged jet multiplicities without applying specific b-tagging selections,
the counting analysis requires at least one b-tagged jet in an accepted event.
This enriches the purity of the \ttbar\ signal in the  data sample, but leads to only 5\% loss in the number of \ttbar\ events.
The values of $N$, $N_B$, and $\mathcal{A}$ are given in Table~\ref{tab:xsecChannels}.

The  cross sections measured with this method are given in Table~\ref{tab:xsec}.
As for the PLR method, the systematic uncertainties are described in Section~\ref{sec:systematics} and summarized in
Tables~\ref{tab:all_sys_max_signal} and~\ref{tab:xsecChannels}.
A breakdown of the statistical and systematic uncertainties contributing to
the combined measurement is given in Table~\ref{tab:breakdown_comb}.
The measurements agree very well with those based on the PLR method.

Compared to the PLR method, the analysis presented in this section follows an alternative approach that leads to a
slightly less sensitive measurement. The PLR and the counting analysis measurements are not combined since they are highly correlated:
both are extracted from almost the same data sample, and are affected by essentially the same systematic uncertainties.

\begin{table}[htb]
\begin{center}
\topcaption{Summary of the individual contributions to the systematic uncertainty on the $\sigma_{\ttbar}$  measurement
for the combined dilepton sample, using the counting analysis. The uncertainties are given in pb.
The statistical uncertainty on the result is given for comparison.}
\begin{tabular}{lc}
\hline
\hline
Source             &  Uncertainty on $\sigma_{\rm {t\bar{t}}}$(pb)  \\
\hline
Diboson       &  0.4 \\
Single top   &  2.3\\
Drell-Yan           &  1.0 \\
Non-W/Z leptons     &  0.6 \\
Lepton efficiencies &  1.7\\
Lepton energy scale &  0.5\\
Jet energy scale &  2.8\\
Jet energy resolution &  0.5\\
\met\ efficiency &  1.9\\
b-tagging  &  1.1\\
Pileup    &  0.7\\
Scale of QCD ($\mu$) &  1.0 \\
Matching partons to showers &  1.0 \\
W branching fraction             &  2.7\\
\hline
Total systematic   &  5.6 \\
\hline
Integrated luminosity         &  3.6 \\
\hline
Statistical         &  2.6 \\
\hline
\hline
\end{tabular}
\label{tab:breakdown_comb}
\end{center}
\end{table}

\subsection{Cross section and mass of the top quark}
\label{sec:summary}
As the acceptance for \ttbar\ signal depends on the top-quark mass, the measured \ttbar\ cross section also depends on the value of the
top-quark mass used to simulate \ttbar\ events. The results presented in Table~\ref{tab:xsec}
for both the PLR and the counting analysis use a top-quark mass of 172.5\GeV
without any additional uncertainties attributed to this quantity.

The dependence of the cross section on  $m_{\cPqt}$ in the range 160--185\GeV is studied by measuring $\sigma_{\ttbar}$ in
\ttbar\ samples simulated at different $m_{\cPqt}$ values. The cross-section dependence can be parametrized as
\begin{equation}
\sigma_{\ttbar}/\sigma_{\ttbar}(m_{\cPqt}  = 172.5) \;=\; 1.00 - 0.008\times (m_{\cPqt}-172.5) - 0.000137 \times (m_{\cPqt}-172.5)^2 .
\end{equation}

For completeness, the combined cross section is also extracted using the PLR method at the currently accepted value $m_{\rm t}$ = $172.9\pm0.6\,\text{(stat.)}\pm0.9\,\text{(syst.)}\GeV$~\cite{PDG2012},
which yields  a cross section of $\sigma_{\ttbar} = 161.3 \pm 2.5\,\text{(stat.)}\,{}^{+5.3}_{-5.2}\,\text{(syst.)}\pm 3.6\,\text{(lumi.)}\unit{pb}$,
where the systematic uncertainty also accounts for the uncertainty on the value of $m_{\cPqt}$.

\section{Summary}
\label{sec:conclusions}
A measurement of the \ttbar production cross section at $\sqrt{s} =7\TeV$ is presented for events containing \eepm, \mmpm, or \empm\ lepton pairs,
at least two jets, and a large imbalance in  transverse momentum. The measurement is performed using two approaches: a profile likelihood-ratio procedure and an event-counting
analysis that relies on the presence of at least one b-tagged jet. The results from the individual dilepton channels and from the two analysis methods
are consistent with each other, and are also found to be compatible with previous published measurements.
For the profile likelihood-ratio method, the cross section for the combined dilepton channels amounts to $\sigma_{\ttbar}=161.9\pm2.5\,\text{(stat.)}\,{}^{+5.1}_{-5.0}\,\text{(syst.)}\pm3.6\,\text{(lumi.)}\unit{pb}$, in agreement with the prediction of the
 standard model. This result corresponds to the most precise measurement of $\sigma_{\ttbar}$ at 7\TeV.

\section*{Acknowledgments}
\hyphenation{Bundes-ministerium Forschungs-gemeinschaft Forschungs-zentren}
We congratulate our colleagues in the CERN accelerator departments for the excellent performance of the LHC machine,
and thank the technical and administrative staffs at CERN and other CMS institutes for their contributions.
This work was supported by the Austrian Federal Ministry of Science and Research; the Belgium Fonds de la
Recherche Scientifique, and Fonds voor Wetenschappelijk Onderzoek; the Brazilian Funding Agencies
(CNPq, CAPES, FAPERJ, and FAPESP); the Bulgarian Ministry of Education and Science; CERN;
the Chinese Academy of Sciences, Ministry of Science and Technology, and National Natural Science Foundation of China;
the Colombian Funding Agency (COLCIENCIAS); the Croatian Ministry of Science, Education and Sport;
the Research Promotion Foundation, Cyprus; the Estonian Academy of Sciences and NICPB;
the Academy of Finland, Finnish Ministry of Education and Culture, and Helsinki Institute of Physics;
the Institut National de Physique Nucl\'eaire et de Physique des Particules~/~CNRS,
and Commissariat \`a l'\'Energie Atomique et aux \'Energies Alternatives~/~CEA, France;
the Bundesministerium f\"ur Bildung und Forschung, Deutsche Forschungsgemeinschaft,
and Helmholtz-Gemeinschaft Deutscher Forschungszentren, Germany; the General Secretariat for Research and Technology, Greece;
the National Scientific Research Foundation, and National Office for Research and Technology, Hungary;
the Department of Atomic Energy and the Department of Science and Technology, India;
the Institute for Studies in Theoretical Physics and Mathematics, Iran; the Science Foundation, Ireland;
the Istituto Nazionale di Fisica Nucleare, Italy;
the Korean Ministry of Education, Science and Technology and the World Class University program of NRF, Korea;
the Lithuanian Academy of Sciences; the Mexican Funding Agencies (CINVESTAV, CONACYT, SEP, and UASLP-FAI);
the Ministry of Science and Innovation, New Zealand; the Pakistan Atomic Energy Commission;
the State Commission for Scientific Research, Poland; the Funda\c{c}\~ao para a Ci\^encia e a Tecnologia, Portugal;
JINR (Armenia, Belarus, Georgia, Ukraine, Uzbekistan);
the Ministry of Science and Technologies of the Russian Federation, and Russian Ministry of Atomic Energy;
the Ministry of Science and Technological Development of Serbia;
the Ministerio de Ciencia e Innovaci\'on, and Programa Consolider-Ingenio 2010, Spain;
the Swiss Funding Agencies (ETH Board, ETH Zurich, PSI, SNF, UniZH, Canton Zurich, and SER); the National Science Council, Taipei;
the Scientific and Technical Research Council of Turkey, and Turkish Atomic Energy Authority;
the Science and Technology Facilities Council, UK; the US Department of Energy, and the US National Science Foundation.

Individuals have received support from the Marie-Curie programme and the European Research Council (European Union);
the Leventis Foundation; the A. P. Sloan Foundation; the Alexander von Humboldt Foundation;
the Associazione per lo Sviluppo Scientifico e Tecnologico del Piemonte (Italy); the Belgian Federal Science Policy Office;
the Fonds pour la Formation \`a la Recherche dans l'Industrie et dans l'Agriculture (FRIA-Belgium);
the Agentschap voor Innovatie door Wetenschap en Technologie (IWT-Belgium); and the Council of Science and Industrial Research, India.

\bibliography{auto_generated}

\providecommand{\href}[2]{#2}\begingroup\raggedright\begin{thebibliography}{10}%
\makeatletter
\providecommand{\hrefCMSnoop }[0]{\@secondoftwo}%
\makeatother
\providecommand{\doi}{\texttt{doi:}\begingroup \urlstyle{tt}\Url}

\bibitem{top10001}
\hrefCMSnoop {} {{ CMS} Collaboration, ``{First Measurement of the Cross
  Section for Top-Quark Pair Production in Proton-Proton Collisions at
  $\sqrt{s}$ = 7 TeV}'',} \textit{ Phys. Lett. B} \textbf{ 695} (2011) 424,
  \href{http://dx.doi.org/10.1016/j.physletb.2010.11.058}{\doi{10.1016/j.physletb.2010.11.058}},
\href{http://www.arXiv.org/abs/1010.5994}{\texttt{ arXiv:1010.5994}}.
%%CITATION = 1010.5994;%%.

\bibitem{ATLAStopPublication4}
\hrefCMSnoop {} {{ ATLAS} Collaboration, ``Measurement of the top quark-pair
  production cross section with ATLAS in pp collisions at $\sqrt{s}$ = 7
  TeV'',} \textit{ Eur. Phys. J. C} \textbf{ 71} (2011) 1577,
  \href{http://dx.doi.org/10.1140/epjc/s10052-011-1577-6}{\doi{10.1140/epjc/s10052-011-1577-6}},
  \href{http://www.arXiv.org/abs/1012.1792}{\texttt{ arXiv:1012.1792}}.

\bibitem{ATLAStopPublication5}
\hrefCMSnoop {} {{ ATLAS} Collaboration, ``Measurement of the cross section for
  top-quark pair production in pp collisions at $\sqrt{s}$ = 7 TeV with the
  ATLAS detector using final states with two high-pt leptons'',} \textit{ JHEP}
  \textbf{ 1205} (2012) 059,
  \href{http://dx.doi.org/10.1007/JHEP05(2012)059}{\doi{10.1007/JHEP05(2012)059}},
  \href{http://www.arXiv.org/abs/1202.4892}{\texttt{ arXiv:1202.4892}}.

\bibitem{ATLAStopPublication6}
\hrefCMSnoop {} {{ ATLAS} Collaboration, ``Measurement of the top quark pair
  production cross-section with ATLAS in the single lepton channel'',} \textit{
  Phys. Lett. B.} \textbf{ 711} (2012) 244,
  \href{http://dx.doi.org/10.1016/j.physletb.2012.03.083}{\doi{10.1016/j.physletb.2012.03.083}},
  \href{http://www.arXiv.org/abs/1201.1889}{\texttt{ arXiv:1201.1889}}.

\bibitem{CMStopPublication2}
\hrefCMSnoop {} {{ CMS} Collaboration, ``Measurement of the $\rm{t\bar{t}}$
  production cross section and the top quark mass in the dilepton channel in pp
  collisions at $\sqrt{s}$ = 7 TeV'',} \textit{ JHEP} \textbf{ 07} (2011) 049,
  \href{http://dx.doi.org/10.1007/JHEP07(2011)049}{\doi{10.1007/JHEP07(2011)049}},
  \href{http://www.arXiv.org/abs/1105.5661}{\texttt{ arXiv:1105.5661}}.

\bibitem{CMStopPublication3}
\hrefCMSnoop {} {{ CMS} Collaboration, ``Measurement of the $\rm{t\bar{t}}$
  Production cross section in pp Collisions at $\sqrt{s}$ = 7 TeV using the
  Kinematic Properties of Events with Leptons and Jets'',} \textit{ Eur. Phys.
  J. C} \textbf{ 71} (2011) 1721,
  \href{http://dx.doi.org/10.1140/epjc/s10052-011-1721-3}{\doi{10.1140/epjc/s10052-011-1721-3}},
  \href{http://www.arXiv.org/abs/1106.0902}{\texttt{ arXiv:1106.0902}}.

\bibitem{CMStopPublication5}
\hrefCMSnoop {} {{ CMS} Collaboration, ``Measurement of the $\rm{t\bar{t}}$
  Production cross section in pp Collisions at 7 TeV in Lepton + Jets Events
  using b-quark Jet Identification'',} \textit{ Phys. Rev. D} \textbf{ 84}
  (2011) 092004,
  \href{http://dx.doi.org/10.1103/PhysRevD.84.092004}{\doi{10.1103/PhysRevD.84.092004}},
  \href{http://www.arXiv.org/abs/1108.3773}{\texttt{ arXiv:1108.3773}}.

\bibitem{ATLASttWPublication}
\hrefCMSnoop {} {{ ATLAS} Collaboration, ``Evidence for the associated
  production of a W boson and a top quark in ATLAS at $\sqrt{s}$ = 7 TeV'',}
  (2012). \href{http://www.arXiv.org/abs/1205.5764}{\texttt{ arXiv:1205.5764}}.
  Submitted to Phys. Lett. B.

\bibitem{CMStopPublication4}
\hrefCMSnoop {} {{ CMS} Collaboration, ``Measurement of the t-channel single
  top quark production cross section in pp collisions at $\sqrt{s}$ = 7 TeV'',}
  \textit{ Phys. Rev. Lett.} \textbf{ 107} (2011) 091802,
  \href{http://dx.doi.org/10.1103/PhysRevLett.107.091802}{\doi{10.1103/PhysRevLett.107.091802}},
  \href{http://www.arXiv.org/abs/1106.3052}{\texttt{ arXiv:1106.3052}}.

\bibitem{ATLASttchenlPublication}
\hrefCMSnoop {} {{ ATLAS} Collaboration, ``{Measurement of the t-channel single
  top-quark production cross section in pp collisions at $\sqrt({s} = 7$ TeV
  with the ATLAS detector}'',} (2012).
  \href{http://www.arXiv.org/abs/1205.3130}{\texttt{ arXiv:1205.3130}}.
  Submitted Phys. Lett. B.

\bibitem{CMStopPublication6}
\hrefCMSnoop {} {{ CMS} Collaboration, ``Measurement of the charge asymmetry in
  top-quark pair production in proton-proton collisions at $\sqrt{s}$ = 7
  TeV'',} \textit{ Phys. Lett. B} \textbf{ 709} (2012) 28,
  \href{http://dx.doi.org/10.1016/j.physletb.2012.01.078}{\doi{10.1016/j.physletb.2012.01.078}},
  \href{http://www.arXiv.org/abs/1112.5100}{\texttt{ arXiv:1112.5100}}.

\bibitem{ATLAStopPublication1}
\hrefCMSnoop {} {{ ATLAS} Collaboration, ``Search for anomalous production of
  prompt like-sign muon pairs and constraints on physics beyond the Standard
  Model'',} \textit{ Phys. Rev. D} \textbf{ 88} (2012) 032004,
  \href{http://dx.doi.org/10.1103/PhysRevD.85.032004}{\doi{10.1103/PhysRevD.85.032004}},
  \href{http://www.arXiv.org/abs/1201.1091}{\texttt{ arXiv:1201.1091}}.

\bibitem{ATLAStopPublication2}
\hrefCMSnoop {} {{ ATLAS} Collaboration, ``Search for new phenomena in
  $\rm{t\bar{t}}$ events with large missing transverse momentum'',} \textit{
  Phys. Rev. Lett.} \textbf{ 108} (2012) 041805,
  \href{http://dx.doi.org/10.1103/PhysRevLett.108.041805}{\doi{10.1103/PhysRevLett.108.041805}},
  \href{http://www.arXiv.org/abs/1109.4725}{\texttt{ arXiv:1109.4725}}.

\bibitem{JINST}
\hrefCMSnoop {} {{ CMS} Collaboration, ``The {CMS} experiment at the {CERN}
  {LHC}'',} \textit{ JINST} \textbf{ 3} (2008) S08004,
\href{http://dx.doi.org/10.1088/1748-0221/3/08/S08004}{\doi{10.1088/1748-0221/3/08/S08004}}.
%%CITATION = JINST,3,S08004;%%.

\bibitem{Xsec_taus}
\hrefCMSnoop {} {{ CMS} Collaboration, ``First measurement of the top quark
  pair production cross section in the dilepton channel with tau leptons in the
  final state in pp collisions at $\sqrt{s}$ = 7 TeV'',} \textit{ Phys.Rev.D}
  \textbf{ 85} (2012) 112007,
  \href{http://dx.doi.org/10.1103/PhysRevD.85.112007}{\doi{10.1103/PhysRevD.85.112007}},
  \href{http://www.arXiv.org/abs/1203.6810}{\texttt{ arXiv:1203.6810}}.

\bibitem{PDG2012}
\hrefCMSnoop {} {{ Particle Data Group} Collaboration, ``{The Review of
  Particle Physics}'',} \textit{ Phys. Rev. D} \textbf{ 86} (2012) 010001,
  \href{http://dx.doi.org/10.1103/PhysRevD.86.010001}{\doi{10.1103/PhysRevD.86.010001}}.

\bibitem{plr1}
A.~Stuart, K.~Ord, and S.~Arnold, ``Kendall's advanced theory of statistics.
  Volume 2A: classical inference and the linear model''.
\newblock Hodder Arnold, 1999.

\bibitem{plr2}
\hrefCMSnoop {} {G.~Cowan {et~al.}, ``Asymptotic formulae for likelihood-based
  tests of new physics'',} \textit{ Eur. Phys. J. C} \textbf{ 71} (2011) 1554,
  \href{http://dx.doi.org/10.1140/epjc/s10052-011-1554-0}{\doi{10.1140/epjc/s10052-011-1554-0}},
  \href{http://www.arXiv.org/abs/1007.1727}{\texttt{ arXiv:1007.1727}}.

\bibitem{mcfm}
\hrefCMSnoop {} {J.~M. Campbell and R.~K. Ellis, ``{MCFM for the Tevatron and
  the LHC}'',} \textit{ Nucl. Phys. Proc. Suppl.} \textbf{ 205-206} (2010) 10,
  \href{http://dx.doi.org/10.1016/j.nuclphysbps.2010.08.011}{\doi{10.1016/j.nuclphysbps.2010.08.011}},
  \href{http://www.arXiv.org/abs/1007.3492}{\texttt{ arXiv:1007.3492}}.

\bibitem{mcfm:tt}
\hrefCMSnoop {} {R.~Kleiss and W.~J. Stirling, ``{Top quark production at
  hadron colliders: Some useful formulae}'',} \textit{ Z. Phys. C} \textbf{ 40}
  (1988) 419,
\href{http://dx.doi.org/10.1007/BF01548856}{\doi{10.1007/BF01548856}}.
%%CITATION = ZEPYA,C40,419;%%.

\bibitem{kidonakis:2010dk}
\hrefCMSnoop {} {N.~Kidonakis, ``{Next-to-next-to-leading soft-gluon
  corrections for the top quark cross section and transverse momentum
  distribution}'',} \textit{ Phys. Rev. D} \textbf{ 82} (2010) 114030,
  \href{http://dx.doi.org/10.1103/PhysRevD.82.114030}{\doi{10.1103/PhysRevD.82.114030}},
  \href{http://www.arXiv.org/abs/1009.4935}{\texttt{ arXiv:1009.4935}}.

\bibitem{Ahrens:2010zv}
\hrefCMSnoop {} {V.~Ahrens {et~al.}, ``{Renormalization-Group Improved
  Predictions for Top-Quark Pair Production at Hadron Colliders}'',} \textit{
  JHEP} \textbf{ 09} (2010) 097,
  \href{http://dx.doi.org/10.1007/JHEP09(2010)097}{\doi{10.1007/JHEP09(2010)097}},
\href{http://www.arXiv.org/abs/1003.5827}{\texttt{ arXiv:1003.5827}}.
%%CITATION = 1003.5827;%%.

\bibitem{Langenfeld:20011}
\hrefCMSnoop {} {M.~Aliev {et~al.}, ``{HATHOR: HAdronic Top and Heavy quarks
  cross section calculatoR.}'',} \textit{ Comput. Phys. Commun.} \textbf{ 182}
  (2011) 1034,
  \href{http://dx.doi.org/10.1016/j.cpc.2010.12.040}{\doi{10.1016/j.cpc.2010.12.040}},
  \href{http://www.arXiv.org/abs/1007.1327}{\texttt{ arXiv:1007.1327}}.

\bibitem{madgraph}
J.~Alwall\hrefCMSnoop {} { {et~al.}, ``{MadGraph 5}: going beyond'',} \textit{
  JHEP} \textbf{ 06} (2011) 128,
  \href{http://dx.doi.org/10.1007/JHEP06(2011)128}{\doi{10.1007/JHEP06(2011)128}},
\href{http://www.arXiv.org/abs/1106.0522}{\texttt{ arXiv:1106.0522}}.
%%CITATION = 1106.0522;%%.

\bibitem{powheg}
\hrefCMSnoop {} {S.~Frixione, P.~Nason, and C.~Oleari, ``{Matching NLO QCD
  computations with parton shower simulations: the POWHEG method}'',} \textit{
  JHEP} \textbf{ 11} (2007) 070,
  \href{http://dx.doi.org/10.1088/1126-6708/2007/11/070}{\doi{10.1088/1126-6708/2007/11/070}},
  \href{http://www.arXiv.org/abs/0709.2092}{\texttt{ arXiv:0709.2092}}.

\bibitem{pythia}
\hrefCMSnoop {} {T.~Sj{\"o}strand, S.~Mrenna, and P.~Skands, ``{PYTHIA 6.4
  physics and manual}'',} \textit{ JHEP} \textbf{ 05} (2006) 026,
  \href{http://dx.doi.org/10.1088/1126-6708/2006/05/026}{\doi{10.1088/1126-6708/2006/05/026}},
\href{http://www.arXiv.org/abs/hep-ph/0603175}{\texttt{ arXiv:hep-ph/0603175}}.
%%CITATION = HEP-PH/0603175;%%.

\bibitem{tauola}
\hrefCMSnoop {} {N.~Davidson {et~al.}, ``{Universal Interface of TAUOLA
  Technical and Physics Documentation}'',} \textit{ Computer Physics
  Communications} \textbf{ 183, Issue 3} (2010) 821,
  \href{http://dx.doi.org/10.1016/j.cpc.2011.12.009}{\doi{10.1016/j.cpc.2011.12.009}},
\href{http://www.arXiv.org/abs/1002.0543}{\texttt{ arXiv:1002.0543}}.
%%CITATION = 1002.0543;%%.

\bibitem{fewz}
\hrefCMSnoop {} {K.~Melnikov and F.~Petriello, ``{Electroweak gauge boson
  production at hadron colliders through $O(\alpha_s^2)$}'',} \textit{ Phys.
  Rev. D} \textbf{ 74} (2006) 114017,
  \href{http://dx.doi.org/10.1103/PhysRevD.74.114017}{\doi{10.1103/PhysRevD.74.114017}},
\href{http://www.arXiv.org/abs/hep-ph/0609070}{\texttt{ arXiv:hep-ph/0609070}}.
%%CITATION = HEP-PH/0609070;%%.

\bibitem{kidonakis}
\hrefCMSnoop {} {N.~Kidonakis, ``Two-loop soft anomalous dimensions for single
  top quark associated production with a W$^-$ or H$^-$'',} \textit{ Phys. Rev.
  D} \textbf{ 82} (2010) 054018,
  \href{http://dx.doi.org/10.1103/PhysRevD.82.054018}{\doi{10.1103/PhysRevD.82.054018}},
\href{http://www.arXiv.org/abs/1005.4451}{\texttt{ arXiv:1005.4451}}.
%%CITATION = 1005.4451;%%.

\bibitem{CMStopPublication7}
\hrefCMSnoop {} {{ CMS} Collaboration, ``Measurement of WW Production and
  Search for the Higgs Boson in pp Collisions at $\sqrt{s}$ = 7 TeV'',}
  \textit{ Phys. Lett. B} \textbf{ 699} (2011) 25,
  \href{http://dx.doi.org/10.1016/j.physletb.2011.03.056}{\doi{10.1016/j.physletb.2011.03.056}},
  \href{http://www.arXiv.org/abs/1102.5429}{\texttt{ arXiv:1102.5429}}.

\bibitem{ATLASWWPublication}
\hrefCMSnoop {} {{ ATLAS} Collaboration, ``Measurement of the $WW$ cross
  section in $\sqrt{s}$ = 7 TeV $pp$ collisions with the ATLAS detector and
  limits on anomalous gauge couplings'',} \textit{ Phys. Lett. B} \textbf{ 712}
  (2012) 289,
  \href{http://dx.doi.org/10.1016/j.physletb.2012.05.003}{\doi{10.1016/j.physletb.2012.05.003}},
  \href{http://www.arXiv.org/abs/1203.6232}{\texttt{ arXiv:1203.6232}}.

\bibitem{ATLASWZPublication}
\hrefCMSnoop {} {{ ATLAS} Collaboration, ``Measurement of the W$^{\pm}$Z
  production cross section and limits on anomalous triple gauge couplings in
  proton-proton collisions at $\sqrt{s}$ = 7 TeV with the ATLAS detector'',}
  \textit{ Phys. Lett. B} \textbf{ 709} (2012) 341,
  \href{http://dx.doi.org/10.1016/j.physletb.2012.02.053}{\doi{10.1016/j.physletb.2012.02.053}},
  \href{http://www.arXiv.org/abs/1111.5570}{\texttt{ arXiv:1111.5570}}.

\bibitem{ATLASZZPublication}
\hrefCMSnoop {} {{ ATLAS} Collaboration, ``Measurement of the $ZZ$ Production
  Cross Section and Limits on Anomalous Neutral Triple Gauge Couplings in
  Proton-Proton Collisions at $\sqrt{s}$ = 7 TeV with the ATLAS Detector'',}
  \textit{ Phys. Rev. Lett.} \textbf{ 108} (2012) 041804,
  \href{http://dx.doi.org/10.1103/PhysRevLett.108.041804}{\doi{10.1103/PhysRevLett.108.041804}},
  \href{http://www.arXiv.org/abs/1110.5016}{\texttt{ arXiv:1110.5016}}.

\bibitem{geant}
\hrefCMSnoop {} {J.~Allison {et~al.}, ``{Geant4 developments and
  applications}'',} \textit{ IEEE Trans. Nucl. Sci.} \textbf{ 53} (2006) 270,
\href{http://dx.doi.org/10.1109/TNS.2006.869826}{\doi{10.1109/TNS.2006.869826}}.
%%CITATION = IETNA,53,270;%%.

\bibitem{trkpas}
\hrefCMSnoop {} {{ CMS} Collaboration, ``{CMS Tracking Performance Results from
  Early LHC Operation}'',} \textit{ Eur. Phys. J. C} \textbf{ 70} (2010) 1165,
  \href{http://dx.doi.org/10.1140/epjc/s10052-010-1491-3}{\doi{10.1140/epjc/s10052-010-1491-3}},
  \href{http://www.arXiv.org/abs/1007.1988}{\texttt{ arXiv:1007.1988}}.

\bibitem{EGMPAS}
\href {http://cdsweb.cern.ch/record/1299116} {{ CMS} Collaboration, ``Electron
  Reconstruction and Identification at $\sqrt{s} = 7$ {TeV}'',} CMS Physics
  Analysis Summary CMS-PAS-EGM-10-004, (2010).

\bibitem{MUOPAS}
\hrefCMSnoop {} {{ CMS} Collaboration, ``Performance of CMS muon reconstruction
  in pp collisions at $\sqrt{s}$ = 7 {TeV}'',} (2012).
  \href{http://www.arXiv.org/abs/1206.4071}{\texttt{ arXiv:1206.4071}}.
  Submitted to J. Instrum.

\bibitem{PFPAS}
\href {http://cdsweb.cern.ch/record/1279341} {{ CMS} Collaboration,
  ``Commissioning of the Particle-Flow Reconstruction in Minimum-Bias and Jet
  Events from {\Pp\Pp} Collisions at 7 {TeV}'',} CMS Physics Analysis Summary
  CMS-PAS-PFT-10-002, (2010).

\bibitem{wzPAS2010}
\hrefCMSnoop {} {{ CMS} Collaboration, ``Measurement of the inclusive W and Z
  production cross sections in pp collisions at $\sqrt{s}$ = 7 TeV with the CMS
  experiment'',} \textit{ JHEP} \textbf{ 10} (2011) 007,
  \href{http://dx.doi.org/10.1007/JHEP10(2011)007}{\doi{10.1007/JHEP10(2011)007}},
  \href{http://www.arXiv.org/abs/1108.0566}{\texttt{ arXiv:1108.0566}}.

\bibitem{antikt}
\hrefCMSnoop {} {M.~Cacciari, G.~P. Salam, and G.~Soyez, ``{The anti-$k_t$ jet
  clustering algorithm}'',} \textit{ JHEP} \textbf{ 04} (2008) 063,
  \href{http://dx.doi.org/10.1088/1126-6708/2008/04/063}{\doi{10.1088/1126-6708/2008/04/063}},
\href{http://www.arXiv.org/abs/0802.1189}{\texttt{ arXiv:0802.1189}}.
%%CITATION = 0802.1189;%%.

\bibitem{JETPAS}
\href {http://cdsweb.cern.ch/record/1279362} {{ CMS} Collaboration, ``Jet
  Performance in pp Collisions at $\sqrt{s}$=7 {TeV}'',} CMS Physics Analysis
  Summary CMS-PAS-JME-10-003, (2010).

\bibitem{fastjet1}
\hrefCMSnoop {} {M.~Cacciari, G.~P. Salam, and G.~Soyez, ``{The Catchment Area
  of Jets}'',} \textit{ JHEP} \textbf{ 04} (2008) 005,
  \href{http://dx.doi.org/10.1088/1126-6708/2008/04/05}{\doi{10.1088/1126-6708/2008/04/05}},
  \href{http://www.arXiv.org/abs/0802.1188}{\texttt{ arXiv:0802.1188}}.

\bibitem{fastjet2}
\hrefCMSnoop {} {M.~Cacciari and G.~P. Salam, ``{Pileup subtraction using jet
  areas}'',} \textit{ Phys. Lett. B} \textbf{ 659} (2008) 119,
  \href{http://dx.doi.org/10.1016/j.physletb.2007.09.077}{\doi{10.1016/j.physletb.2007.09.077}},
  \href{http://www.arXiv.org/abs/0707.1378}{\texttt{ arXiv:0707.1378}}.

\bibitem{METPAS2}
\href {http://cdsweb.cern.ch/record/1294501} {{ CMS} Collaboration, ``CMS MET
  Performance in Events Containing Electroweak Bosons from pp Collisions at
  $\sqrt{s}=7$ {TeV}'',} CMS Physics Analysis Summary CMS-PAS-JME-10-005,
  (2010).

\bibitem{BTV-11-004}
\href {http://cdsweb.cern.ch/record/1427247} {{ CMS} Collaboration, ``b-Jet
  Identification in the {CMS} Experiment'',} CMS Physics Analysis Summary
  CMS-PAS-BTV-11-004, (2011).

\bibitem{BTV-11-003}
\href {http://cdsweb.cern.ch/record/1421611} {{ CMS} Collaboration,
  ``Measurement of the b-tagging efficiency using \ttbar\ events'',} CMS
  Physics Analysis Summary CMS-PAS-BTV-11-003, (2011).

\bibitem{lumipas}
\href {http://cdsweb.cern.ch/record/1376102} {{ CMS} Collaboration, ``Absolute
  Calibration of the {CMS} Luminosity Measurement: {S}ummer 2011 Update'',} CMS
  Physics Analysis Summary CMS-PAS-EWK-11-001, (2011).

\bibitem{JESPAS}
\href {http://cdsweb.cern.ch/record/1308178} {{ CMS} Collaboration,
  ``Determination of the Jet Energy Scale in {CMS} with pp Collisions at
  $\sqrt{s}= 7$ {TeV}'',} CMS Physics Analysis Summary CMS-PAS-JME-10-010,
  (2010).

\bibitem{TOTEM73.5mb}
\hrefCMSnoop {} {{ TOTEM} Collaboration, ``{First measurement of the total
  proton-proton cross section at the LHC energy of $\sqrt{s}=7$ TeV}'',}
  \textit{ Europhys. Lett.} \textbf{ 96} (2011) 21002,
  \href{http://dx.doi.org/10.1209/0295-5075/96/21002}{\doi{10.1209/0295-5075/96/21002}},
\href{http://www.arXiv.org/abs/1110.1395}{\texttt{ arXiv:1110.1395}}.
%%CITATION = 1110.1395;%%.

\bibitem{BLUE}
\hrefCMSnoop {} {L.~Lyons, D.~Gibaut, and P.~Clifford, ``{How to combine
  correlated estimates of a single physical quantity}'',} \textit{ Nucl.
  Instrum. Meth. A} \textbf{ 270} (1988) 110,
\href{http://dx.doi.org/10.1016/0168-9002(88)90018-6}{\doi{10.1016/0168-9002(88)90018-6}}.
%%CITATION = NUIMA,A270,110;%%.

\end{thebibliography}\endgroup

\cleardoublepage \appendix\section{The CMS Collaboration \label{app:collab}}\begin{sloppypar}\hyphenpenalty=5000\widowpenalty=500\clubpenalty=5000\textbf{Yerevan Physics Institute,  Yerevan,  Armenia}\\*[0pt]
S.~Chatrchyan, V.~Khachatryan, A.M.~Sirunyan, A.~Tumasyan
\vskip\cmsinstskip
\textbf{Institut f\"{u}r Hochenergiephysik der OeAW,  Wien,  Austria}\\*[0pt]
W.~Adam, E.~Aguilo, T.~Bergauer, M.~Dragicevic, J.~Er\"{o}, C.~Fabjan\cmsAuthorMark{1}, M.~Friedl, R.~Fr\"{u}hwirth\cmsAuthorMark{1}, V.M.~Ghete, J.~Hammer, N.~H\"{o}rmann, J.~Hrubec, M.~Jeitler\cmsAuthorMark{1}, W.~Kiesenhofer, V.~Kn\"{u}nz, M.~Krammer\cmsAuthorMark{1}, I.~Kr\"{a}tschmer, D.~Liko, I.~Mikulec, M.~Pernicka$^{\textrm{\dag}}$, B.~Rahbaran, C.~Rohringer, H.~Rohringer, R.~Sch\"{o}fbeck, J.~Strauss, A.~Taurok, W.~Waltenberger, G.~Walzel, E.~Widl, C.-E.~Wulz\cmsAuthorMark{1}
% ...
\vskip\cmsinstskip
\dag:~Deceased\\
1:~~Also at Vienna University of Technology, Vienna, Austria\\
% ...

\end{sloppypar}
\end{document}